\documentstyle[preprint,prc,aps,eqsecnum,epsf]{revtex}
\begin{document}
\draft

\title{Photo- and Electro-Disintegration of $^3$He
at Threshold and $p$$d$ Radiative Capture}
\author{M.\ Viviani}
\address{INFN, Sezione di Pisa, I-56100 Pisa, Italy}
\author{A.\ Kievsky}
\address{INFN, Sezione di Pisa, I-56100 Pisa, Italy}
\author{L.\ E.\ Marcucci}
\address{Department of Physics, Old Dominion University,
         Norfolk, Virginia 23529}
\author{S.\ Rosati}
\address{Department of Physics, University of Pisa, I-56100 Pisa, Italy\\
and\\
INFN, Sezione di Pisa, I-56100 Pisa, Italy}
\author{R.\ Schiavilla}
\address{Jefferson Lab, Newport News, Virginia 23606 \\
         and \\
         Department of Physics, Old Dominion University,
         Norfolk, Virginia 23529}
\date{\today}

\maketitle

\begin{abstract}
The present work reports results for: i) $pd$ radiative capture
observables measured at center-of-mass
(c.m.) energies in the range 0--100 keV and
at 2 MeV by the TUNL and Wisconsin groups, respectively;
ii) contributions to the Gerasimov-Drell-Hearn (GDH) integral in $^3$He
from the two- up to the three-body breakup thresholds, compared
to experimental determinations by the TUNL group in this threshold
region; iii) longitudinal, transverse, and interference response
functions measured in inclusive polarized electron scattering
off polarized $^3$He at excitation energies below the threshold
for breakup into $ppn$, compared to unpolarized longitudinal
and transverse data from the Saskatoon group.  The calculations
are based on pair-correlated-hyperspherical-harmonics bound and
continuum wave functions obtained from a realistic
Hamiltonian consisting of the Argonne $v_{18}$ two-nucleon and
Urbana IX three-nucleon interactions.  The electromagnetic current
operator includes one- and two-body components, leading terms
of which are constructed from the Argonne $v_{18}$ interaction
(specifically, its charge-independent part).  Two-body currents associated
with $\Delta$-isobar degrees of freedom are treated non-perturbatively
via the transition-correlation-operator method.  The theoretical
predictions obtained by including only one-body currents are in violent
disagreement with data.  These differences between theory
and experiment are, to a large extent, removed when two-body
currents are taken into account, although some rather large discrepancies remain
in the c.m. energy range 0--100 keV, particularly for the $pd$ differential
cross section $\sigma(\theta)$ and tensor analyzing power $T_{20}(\theta)$
at small angles, and contributions to the GDH integral.  A rather
detailed analysis indicates that these discrepancies have, in large part, a common
origin, and can be traced back to an excess strength obtained
in the theoretical calculation of the $E_1$ reduced
matrix element associated with the $pd$ channel having
$L,S,J=1,1/2,3/2$.  It is 
suggested that this lack of $E_1$ strength observed experimentally
might have implications for the nuclear interaction at very low
energies.  Finally, the validity of the long-wavelength approximation
for electric dipole transitions is discussed.
\end{abstract}
\bigskip

\section{Introduction}
\label{sec:intro}

Radiative capture, photo- and electro-disintegration reactions are a useful
tool for exploring the structure of nuclei and their electromagnetic
response.  The theoretical description of these processes requires
knowledge of the nuclear bound- and scattering-state wave functions and
electromagnetic transition operators.  In this respect, the trinucleon
systems play a unique role because of the capability, achieved in
the last few years, to obtain very accurate wave functions
for both bound and continuum states from realistic Hamiltonian
models~\cite{Bochum,KRV94,KRV95,Pea97}.

The accuracy of the calculated trinucleon continuum wave functions has been
verified by comparing results for a variety of $Nd$ scattering observables
obtained by a number of groups using different techniques (see
Ref.~\cite{Bochum} and references therein).
In fact, good overall agreement exists between theory and experiment for
both elastic and inelastic $Nd$ cross sections and polarization
observables, with the only notable exceptions of the $pd$ and $nd$
vector analyzing powers at low energies, which are both underpredicted
by theory at the 30\% level~\cite{Bochum,KRTV96}.  Indeed, the $A_y$
\lq\lq puzzle\rq\rq\ constitutes an excellent example of how, once the
numerical uncertainties in 
the calculation of the continuum wave functions have been drastically
reduced, $Nd$ scattering observables can be used to study the sentitivity
to two- and three-nucleon interaction models.
 
Electromagnetic processes, such as the low-energy $pd$ radiative fusion
and threshold photo- and electro-disintegration of $^3$He under consideration
in the present work, provide additional
and essential insights into the structure 
of the trinucleons, since they can be
used to test and refine (or possibly, 
discriminate among) models for the
nuclear interactions and electromagnetic transition
operators.  Furthermore, they
allow us to study a number of other, closely
related issues.  A specific example of these
is the relevance of $\Delta$-isobar degrees
of freedom for the proper description of photo- and electro-nuclear
observables~\cite{Viv96}.  

There are now available many high-quality data, including differential
cross sections, vector and tensor analyzing powers, and photon polarization
coefficients, on the $pd$ radiative capture at c.m. energies
ranging from 0 to 2 MeV~\cite{Sea96,Mea97,SK99}.  The goal of the
present study is to 
determine the extent to which this rich body of data can be described
satisfactorily by a calculation based on a realistic Hamiltonian (consisting
of the Argonne $v_{18}$ two-nucleon~\cite{WSS95}
and Urbana-IX three-nucleon~\cite{PPCW95}
interactions), and a current operator including one- and two-body
components, leading terms of which are constructed consistently
with the two-nucleon interaction.  The present work updates
and extends previous ones, which only dealt with
$pd$ radiative capture in the c.m. energy range 0--100 keV and 
$nd$ radiative capture at thermal neutron energies~\cite{Viv96}.
It contains: i) an improved treatment of the $pd$ continuum
within the pair-correlated-hyperspherical-harmonics
(PHH) scheme; ii) calculations of $pd$ radiative capture
polarization observables in the c.m. energy range 0--100 keV,
measured by the TUNL group~\cite{Sea96}, and at 2 MeV,
measured by the Wisconsin group~\cite{SK99};
iii) a calculation of the contribution to the Gerasimov-Drell-Hearn
integral of $^3$He in the threshold region, compared to results of an
experimental determination by the TUNL group~\cite{Wea99}; and iv) a
calculation 
of the longitudinal, transverse, and interference response functions
measured in inclusive scattering of polarized electrons off polarized $^3$He
for excitation energies below the three-body breakup threshold and
for momentum transfers up to 5 fm$^{-1}$, compared with longitudinal
and transverse data measured by Retzlaff {\it et al.}~\cite{Rea94}
at $q$=0.88, 1.54 and 2.47 fm$^{-1}$.

The paper is organized into five sections and an appendix.  In Sec.~II
we discuss the calculation of the bound and scattering wave functions
with the PHH method, and summarize a number of results obtained
for $Nd$ elastic scattering observables, comparing them to experimental data.
In Sec.~III we briefly review the model
for the electromagnetic transition operator, while in Sec.~IV we present
an exhaustive comparison between theory and experiment for all available
$pd$ capture, photo- and electro-disintegration data in the threshold region.
Finally, in Sec.~V we summarize our conclusions.  A collection of
formulas for the calculation of various observables from the
reduced matrix elements of the current and charge operators, which
complement those published earlier in Ref.~\cite{Viv96},
are given in the Appendix. 

\section{Bound- and Scattering-State Wave Functions}
\label{sec:phh}

The $^3$He bound-state and $pd$ scattering-state wave functions
are obtained variationally with the pair-correlated-hyperspherical-harmonics
(PHH) method from a realistic Hamiltonian including
the Argonne $v_{18}$ two-nucleon~\cite{WSS95} and Urbana-IX
three-nucleon~\cite{PPCW95} 
interactions (the AV18/UIX model).  The PHH method, as implemented
in the calculations reported in the present work,
has been developed by Kievsky, Rosati and Viviani
in a series of papers
appeared in various journals between 1993 and 1995~\cite{KRV94,KRV95,KRV93}.
Here it will be reviewed briefly for completeness, and a summary of
relevant results obtained with it for the three-nucleon bound-state
properties and $N$$d$ scattering observables at energies below the three-body
breakup thresholds will be presented.  

The three-nucleon bound-state wave function $\Psi_3$ is expanded as~\cite{KRV93}

\begin{equation}
\Psi_3=\sum_{\alpha, K} \frac{u_{\alpha K}(\rho)}{\rho^{5/2}}
\sum_{{\rm cyclic} \> ijk} Z_{\alpha K}(i;jk) \>\>,
\label{psi3}
\end{equation}
where $\rho$ is the hyperradius, $\rho=\sqrt{x_i^2+y_i^2}$ with
the Jacobi variables ${\bf x}_i$ and ${\bf y}_i$
defined, respectively, as ${\bf x}_i={\bf r}_j-{\bf r}_k$ and
${\bf y}_i=( {\bf r}_j+{\bf r}_k -2\,{\bf r}_i )/\sqrt{3}$.  The
known functions $Z_{\alpha K}(i;jk)$ 
are antisymmetric under the
exchange $j \rightleftharpoons k$ and account for the angle-spin-isospin
and hyperangle dependence of channel $\alpha,K$.  The index $\alpha$
denotes collectively the spectator $i$ and pair $jk$ orbital and spin
angular momenta and isospins coupled to produce
a state with total angular momentum and isospin
$J^\pi,T={\frac{1}{2}}^+,\frac{1}{2}$, while the index $K$ specifies 
the order of the Jacobi
polynomial in the hyperangle ${\rm cos}\, \phi_i = x_i/\rho$. 
Correlation factors, which account for the strong state-dependent correlations 
induced by the nucleon-nucleon interaction, are included 
in the functions $Z_{\alpha K}(i;jk)$.

The Rayleigh-Ritz variational principle,

\begin{equation}
\langle \delta_u \Psi_3 | H-E_3 | \Psi_3\rangle = 0 \>\>\>,
\end{equation}
is used to determine the functions $u_{\alpha K}(\rho)$.  Carrying
out the variations with respect to the $u_{\alpha K}$'s 
leads to a set of coupled second-order differential equations.  After
discretization in the variable $\rho$, this set of differential
equations is converted into a generalized eigenvalue problem, which
is then solved by standard numerical techniques~\cite{KRV93}.

Fully converged AV18/UIX PHH wave functions predict three-nucleon
binding energies and matter radii respectively given by $B(^3{\rm H})=$8.48 MeV
and $\langle r^2(^3{\rm H})\rangle^{1/2} = 1.725$ fm, 
$B(^3{\rm He})=7.73$ MeV and
$\langle r^2(^3{\rm He})\rangle^{1/2} =1.928$ fm,
in agreement with corresponding \lq\lq exact\rq\rq\ Green's function
Monte Carlo results~\cite{Pea97} and with the available experimental
data. 

The $N$$d$ cluster wave function $\Psi_{1+2}^{LSJJ_z}$, having incoming orbital angular
momentum $L$ and channel spin $S$ ($S=1/2,3/2$)
coupled to total $JJ_z$, is expressed as 
\begin{equation}
   \Psi_{1+2}^{LSJJ_z}  = \Psi_{C}^{JJ_z}
   + \Psi_{A}^{LSJJ_z} \ ,\label{eq:scatte}
\end{equation}
where $\Psi_C$ vanishes in the limit of large
intercluster separation, and hence describes the system
in the region where the particles are close to each other and
their mutual interactions are large.  In the asymptotic
region, where intercluster interactions are negligible,
$\Psi_A^{LSJJ_z}$ (for $p$$d$, as an example) is written as

\begin{eqnarray}
   \Psi_{A}^{LSJJ_z}   &=& \sum_{{\rm cyclic}\> ijk} \sum_{L^\prime
   S^\prime} \Big[ \lbrack s_i \otimes \phi_d({\bf x}_i)
   \rbrack_{S^\prime} \otimes Y_{L^\prime}({\hat {\bf y}}_i) \Big]_{JJ_z}
   \nonumber \\
   &&\times \Bigg\lbrack \delta_{L L^\prime} \delta_{S S^\prime}
   { F_{L^\prime}(pr_{pd}) \over pr_{pd} }
    + R^J_{LS,L^\prime S^\prime}(p)
   { G_{L^\prime}(pr_{pd}) \over pr_{pd} }
   g (r_{pd}) \Bigg\rbrack
   \>\>\>, \label{eq:psia}
\end{eqnarray}
where $\phi_d$ is the deuteron wave function, $p$ the
magnitude of the relative momentum between
deuteron and proton, and $F_L$ and $G_L$ are
the regular and irregular Coulomb functions, respectively.  Note that
for $n$$d$ scattering, $F_L(x)/x$ and $G_L(x)/x$
are to be replaced by the
regular and irregular spherical Bessel functions.  The function
$g(r_{pd})$ modifies the $G_L(pr_{pd})$
at small $r_{pd}$ by regularizing
it at the origin, and $g(r_{pd})\rightarrow 1$
as $r_{pd} \geq 10$ fm, thus not affecting
the asymptotic behavior of
$\Psi_{1+2}^{LSJJ_z}$.  Finally, the real parameters $R^J_{LS,L^\prime
S^\prime}(p)$ are the $R$-matrix elements which determine phase-shifts
and (for coupled channels) mixing angles at the energy $p^2/(2\mu)$,
$\mu$ being the 1+2 reduced mass.  Of course, the sum over $L^\prime
S^\prime$ is over all values compatible with a given $J$ and parity.

The \lq\lq core\rq\rq\ wave function $\Psi_C$ is expanded in the same PHH
basis as the bound-state wave function $\Psi_3$, Eq.~(\ref{psi3}),
and both the matrix elements $R^J_{LS,L^\prime S^\prime}(p)$ and
functions $u_{\alpha K}(\rho)$ occurring in the expansion of $\Psi_C$
are determined by making the functional~\cite{KRV94,KRV95}

\begin{equation}
   [R^J_{LS,L^\prime S^\prime}(p)]= R^J_{LS,L^\prime S^\prime}(p)
   -\langle \Psi_{1+2}^{L^\prime S^\prime JJ_z} | H-E_2-\frac{p^2}{2\mu}
   | \Psi_{1+2}^{LSJJ_z} \rangle\ , \label{eq:kohn}
\end{equation}
stationary with respect to variations in the
$R^J_{LS,L^\prime S^\prime}$ and
$u_{\alpha K}$ (Kohn variational principle).  Here $E_2=-2.225$ MeV
is the deuteron energy.   

Phase-shifts and mixing angles for $nd$ scattering at energies
below the three-body breakup threshold have been obtained from
a realistic Hamiltonian model, and have been shown to
be in excellent agreement with corresponding Faddeev results~\cite{Kea98}, thus
establishing the high accuracy of the PHH expansion for the scattering
problem.  It is important to emphasize that the PHH scheme, in contrast
to momentum-space Faddeev methods, permits the straightforward
inclusion of Coulomb distortion effects in the $pd$ channel.
The PHH results for $pd$ elastic
scattering are as accurate as those for $nd$ scattering.

The $nd$ and $pd$ doublet and quartet scattering lengths
predicted by the AV18/UIX model are listed in Table~\ref{tab:lgths},
and are found to be in excellent agreement with the available experimental
values. 
\section{The Electromagnetic Current Operator}
\label{sec:MEC}

The nuclear charge and current operators consist of one- and two-body terms
that operate on the nucleon degrees of freedom:

\begin{eqnarray}
\rho({\bf q})&=& \sum_i \rho^{(1)}_i({\bf q}) 
             +\sum_{i<j} \rho^{(2)}_{ij}({\bf q}) \>\>, \label{eq1}\\
{\bf j}({\bf q})&=& \sum_i {\bf j}^{(1)}_i({\bf q}) 
             +\sum_{i<j} {\bf j}^{(2)}_{ij}({\bf q}) \label{eq2} \>\>,
\end{eqnarray}
where ${\bf q}$ is the momentum transfer.

The one-body operators $\rho^{(1)}_i$ and ${\bf j}^{(1)}_i$ have the
standard expressions obtained from a relativistic reduction of the 
covariant single-nucleon current, and are listed below for convenience.
The charge operator is written as
\begin{equation}
\rho^{(1)}_i({\bf q})= \rho^{(1)}_{i,{\rm NR}}({\bf q})+
                       \rho^{(1)}_{i,{\rm RC}}({\bf q}) \>\>, \label{eq6}
\end{equation}
with
\begin{equation}
\rho^{(1)}_{i,{\rm NR}}({\bf q})= \epsilon_i \>
 {\rm e}^{{\rm i}{\bf q}\cdot {\bf r}_i} \label{eq7} \>\>,
\end{equation}
\begin{equation}
\rho^{(1)}_{i,{\rm RC}}({\bf q})= \left 
( \frac {1}{\sqrt{1+q_{\mu}^2/4m^2} }-1\right ) 
\epsilon_i \> {\rm e}^{{\rm i}{\bf q}\cdot {\bf r}_i} 
- {\frac {{\rm i}}{4m^2}} \left ( 2\, \mu_i-\epsilon_i \right ) 
{\bf q} \cdot (\bbox{\sigma}_i \times {\bf p}_i) \> 
{\rm e}^{ {\rm i} {\bf q} \cdot {\bf r}_i } \>\>, 
\label{eq8} 
\end{equation}
where $q_{\mu}^2=q^2-\omega^2$ is the four-momentum transfer, and
$\omega$ is the energy transfer.  The current operator is expressed as 
\begin{equation} 
     {\bf j}^{(1)}_i({\bf q})={\frac {1} {2m}} \epsilon_i \>
   \bigl[ {\bf p}_i\>,\>{\rm e}^{{\rm i} {\bf q} \cdot {\bf r}_i} \bigr ]_+
   -{\frac {{\rm i}} {2m}} \mu_i \> 
     {\bf q} \times \bbox{\sigma}_i \> {\rm e}^{{\rm i} {\bf q} \cdot
   {\bf r}_i}  \label{eq9}\>\>\>,
\end{equation}
where $[ \cdots \, ,\, \cdots ]_+$ denotes the anticommutator.  The following
definitions have been introduced: 

\begin{eqnarray}
\epsilon_i &\equiv& \frac {1}{2}\left [ G_E^S(q_{\mu}^2) +
G_E^V(q_{\mu}^2)\tau_{z,i} \right ] \>\>,\label{eq8a}\\
\mu_i &\equiv& \frac {1}{2}\left [ G_M^S(q_{\mu}^2) + 
G_M^V(q_{\mu}^2)\tau_{z,i} \right ] \label{eq9a} \>\>,
\end{eqnarray}
and ${\bf p}$, $\bbox{\sigma}$, and $\bbox{\tau}$ are
the nucleon's momentum, Pauli spin
and isospin operators, respectively.  The two terms
proportional to $1/m^2$ in $\rho^{(1)}_{i,{\rm RC}}$
are the well known Darwin-Foldy and spin-orbit relativistic
corrections~\cite{Def66,Fri73}, respectively.  

The dipole parametrization is used for the isoscalar ($S$) and
isovector ($V$) combinations of the electric and magnetic nucleon form
factors (including the Galster form
for the electric neutron form factor~\cite{Gal71}).  It is worth emphasizing
that the neutron form factors, particularly
the electric one, are not well known and, therefore, the available
semi-empirical parameterizations for them
differ widely, particularly at high momentum transfers.  Until
this uncertainty in the detailed behavior
of the electromagnetic form factors of the
nucleon is narrowed, quantitative predictions
of electronuclear observables at high
momentum transfers will remain somewhat tentative. 

The most important features of the two-body parts of the
electromagnetic current operator are summarized below.  The reader
is referred to Refs.~\cite{Viv96,Car98} for a derivation and listing
of their explicit expressions.  

\subsection{Two-body current operators}

The two-body current
operator has \lq\lq model-independent\rq\rq and 
\lq\lq model-dependent \rq\rq components, in the classification scheme 
of Riska~\cite{Ris89}.  
The model-independent terms are obtained from the charge-independent part
of the AV18, and by construction~\cite{Ris85} satisfy current conservation 
with this interaction.  
The leading operator is the isovector \lq\lq $\pi$-like \rq\rq current obtained
from the isospin-dependent spin-spin and tensor interactions.  
The latter also generate an isovector \lq\lq $\rho$-like \rq\rq current, while 
additional model-independent isoscalar and isovector currents arise from the  
isospin-independent and isospin-dependent central and momentum-dependent 
interactions.
These currents are short-ranged and numerically far less important than
the $\pi$-like current.

The model-dependent currents are purely transverse
and therefore cannot be directly linked to the underlying two-nucleon interaction.
The present calculation includes the isoscalar $\rho \pi \gamma$ and 
isovector $\omega \pi \gamma$ transition currents as well as the isovector 
current associated with excitation of intermediate $\Delta$-isobar resonances.  
The $\rho \pi \gamma$ and $\omega \pi \gamma$ couplings are known from the 
measured widths of the radiative decays 
$\rho \rightarrow \pi \gamma$~\cite{Ber80} and 
$\omega \rightarrow \pi \gamma$~\cite{Che71}, respectively, while their 
momentum-transfer dependence is modeled using vector-meson-dominance.  
Monopole form factors are introduced at the meson-baryon vertices with 
cutoff values of $\Lambda_\pi$=3.8 fm$^{-1}$ and
$\Lambda_\rho$=$\Lambda_\omega$=6.3 fm$^{-1}$ 
at the $\pi N$$N$, $\rho N$$N$ and $\omega N$$N$ vertices, respectively.

Among the model-dependent currents, those associated with the $\Delta$-isobar
are the most important ones.  In the present calculation, these currents are
treated within the transition-correlation-operator (TCO) scheme developed 
in Ref.~\cite{Sch92}.  In such an approach, the $\Delta$ degrees of freedom
are explicitly included in the nuclear wave functions by writing

\begin{equation}
\Psi_{N+\Delta}=\left[{\cal{S}}\prod_{i<j}\left(1\,+\,U^{TR}_{ij}\right)
\right]\, \Psi \ ,
\label{eq:psiNDtco}
\end{equation}
where $\Psi$ is the purely nucleonic component, $\cal{S}$ is the
symmetrizer and the transition correlations $U^{TR}_{ij}$ are
short-range operators, that convert $NN$ pairs
into $N\Delta$ and $\Delta\Delta$ pairs.  In the present study the $\Psi$
is taken from PHH solutions of the AV18/UIX Hamiltonian with nucleons only
interactions, while the $U^{TR}_{ij}$ is obtained from two-body
bound and low-energy scattering state solutions of the full $N$-$\Delta$
coupled-channel problem. This aspect of the present calculations as well as
the justification for going beyond the traditional
perturbative treatment of $\Delta$ degrees of freedom in nuclei, were
discussed at length in the original work~\cite{Sch92}, and have been
reviewed most recently in Ref.~\cite{Mar98}, making a further review
here unnecessary.

In the TCO approach, both $\gamma N \Delta$ and $\gamma \Delta \Delta$
$M_1$ couplings are considered~\cite{Sch92}.  The values used for these
couplings are, respectively, $\mu_{\gamma N \Delta}=3$ n.m. and
$\mu_{\gamma \Delta \Delta}=4.35$ n.m.  The former is taken
from an analysis of $\gamma N$ data in the $\Delta$-resonance
region~\cite{Car86}, while the latter is obtained
from a soft-photon analysis of pion-proton bremsstrahlung data
near the $\Delta^{++}$ resonance~\cite{Lin91}.  

Electromagnetic observables require evaluation
of matrix elements of the type~\cite{Sch92,Mar98}

\begin{equation}
\frac{ \langle\Psi_{N+\Delta,f}\,|\,{\bf j}\,|\,\Psi_{N+\Delta,i}\rangle }
{[\langle\Psi_{N+\Delta,f}\,|\,\Psi_{N+\Delta,f}\rangle 
 \langle\Psi_{N+\Delta,i}\,|\,\Psi_{N+\Delta,i}\rangle]^{1/2}} \>\>\>,
\label{eq:emo}
\end{equation}
where the wave functions and currents include both
nucleonic and $\Delta$-isobar degrees of freedom.
To evaluate such a matrix element, it is convenient to expand the wave
function $\Psi_{N+\Delta}$ as
\begin{equation}
\Psi_{N+\Delta} = \Psi + \sum_{i<j}U_{ij}^{TR}\Psi +
\ldots \ ,
\label{eq:psind}
\end{equation}
and write the numerator of Eq.~(\ref{eq:emo}), in a schematic notation, as
\begin{equation}
\langle\Psi_{N+\Delta,f}\,|\,{\bf j}\,|\,\Psi_{N+\Delta,i}\rangle\,=\,\langle
\Psi_{f}\,|\,{\bf j}(N\,{\rm only})\,|\,\Psi_{i}\rangle
\,+\,\langle\Psi_{f}\,|\,{\bf j}(\Delta)\,|\,\Psi_{i}\rangle \ ,
\label{eq:schemj}
\end{equation}
where ${\bf j}(N \,{\rm only})$ denotes all one- and two-body contributions to
${\bf j}({\bf q})$ which only involve nucleon degrees of freedom, i.e.,
${\bf j}(N \,{\rm only})\,=\,{\bf j}^{(1)}(N\rightarrow N)\,+\,{\bf
j}^{(2)}(NN\rightarrow NN)$. 
The operator ${\bf j}(\Delta)$ includes terms involving the $\Delta$-isobar
degrees of freedom, associated with the explicit $\Delta$ currents
${\bf j}^{(1)} 
(N\rightarrow\Delta)$, ${\bf j}^{(1)}(\Delta\rightarrow N)$, and ${\bf
j}^{(1)}(\Delta 
\rightarrow\Delta)$, and with the transition operators $U_{ij}^{TR}$.
Of course, 
the presence of $\Delta$ admixtures in the wave functions also influences 
their normalization. 

Finally, we note that the contributions associated
with the $\rho \pi \gamma$, $\omega \pi \gamma$ and
$\Delta$-excitation mechanisms are, at low and moderate
values of momentum transfer ($q \leq 5$ fm$^{-1}$), typically much
smaller than those due to the leading model-independent
$\pi$-like current~\cite{Viv96,Mar98}.  

\subsection{Two-body charge operators}

While the main parts of the two-body currents are linked to the form of the 
two-nucleon interaction through the continuity equation, the most important 
two-body charge operators are model-dependent, and should be considered as 
relativistic corrections.  
Indeed, a consistent calculation of two-body charge effects in nuclei would 
require the inclusion of relativistic effects in both the interaction models 
and nuclear wave functions.  
Such a program is just at its inception for systems with $A\geq 3$. 
There are nevertheless rather clear indications for the relevance of two-body 
charge operators from the failure of the impulse approximation (IA) in 
predicting the deuteron tensor polarization
observable~\cite{T2098}, and charge form factors of the three- and
four-nucleon systems~\cite{Car98,Mar98}.  
The model commonly used~\cite{Sch90} includes the $\pi$-, $\rho$-, and 
$\omega$-meson exchange charge operators with both isoscalar and isovector 
components, as well as the (isoscalar) $\rho \pi \gamma$ and (isovector) 
$\omega \pi \gamma$ charge transition couplings, in addition to the 
single-nucleon Darwin-Foldy and spin-orbit relativistic corrections.  
The $\pi$- and $\rho$-meson exchange charge operators are constructed from the
AV18 isospin-dependent spin-spin and tensor interactions, using the same
prescription adopted for the corresponding current operators.
It should be emphasized, however, that for $q \!< \! 5$ fm$^{-1}$ the 
contribution due to the $\pi$-exchange charge operator is typically an order 
of magnitude larger than that of any of the remaining two-body mechanisms 
and one-body relativistic corrections.

\section{Results}
\label{sec:RES}

The present section contains results for the $pd$ radiative
capture in the c.m. energy range 0--100 keV, compared to 
the TUNL data~\cite{Sea96,Mea97}, and at 2 MeV, compared to the
Wisconsin data~\cite{SK99}, as well as results for the threshold
electrodisintegration of $^3$He at momentum transfers $q$=0.88, 1.64
and 2.74 fm$^{-1}$, compared to Bates data~\cite{Rea94}.  It also
contains predictions for the contribution to the Gerasimov-Drell-Hearn
integral of $^3$He up to the three-body breakup threshold, compared to
a very recent experimental determination of this contribution
in the $pd$ threshold region by the TUNL group~\cite{Wea99}. 

The calculations are based on bound- and scattering-state PHH
wave functions, obtained from
the Argonne $v_{18}$ two-nucleon and Urbana-IX three-nucleon
interactions (see discussion in Sec.~\ref{sec:phh}).  The model
for the electromagnetic current operator includes one- and two-body
components, leading terms of which are constructed from the two-nucleon
interaction (in the present case, the charge independent part
of the AV18).  The currents associated with the excitation
of $\Delta$-isobars are treated non-perturbatively within the
transition-correlation-operator scheme, as sketched in Sec.~\ref{sec:MEC}.

A study of the $pd$ capture cross section and
polarization observables in the c.m. energy range 0--100 keV
was reported earlier in Ref.~\cite{Viv96}.  In the present work,
improvements in the PHH variational treatment of the P-wave $pd$
channel have led to significant changes in the predictions for the
$S$-factor and some of the polarization observables, particularly
the vector analyzing power $A_y$, than previously published~\cite{Viv96}.
We therefore provide an update of that study.

\subsection{The $pd$ capture below 100 keV c.m. energy}

The observed linear dependence upon the energy of the
$S$-factor and the observed angular distributions of the polarization
observables indicate that the $pd$ radiative fusion proceeds, at these
low energies,  through S- and P-wave captures.  Therefore, the contributing
reduced matrix elements (RMEs) are
$M_1^{0\frac{1}{2}\frac{1}{2}}$, $M_1^{0\frac{3}{2}\frac{3}{2}}$, and 
$E_2^{0\frac{3}{2}\frac{3}{2}}$ in S-wave capture, and
$E_1^{1\frac{1}{2}\frac{1}{2}}$, $E_1^{1\frac{1}{2}\frac{3}{2}}$, 
$E_1^{1\frac{3}{2}\frac{1}{2}}$, $E_1^{1\frac{3}{2}\frac{3}{2}}$, 
$M_2^{1\frac{1}{2}\frac{3}{2}}$, and $M_2^{1\frac{3}{2}\frac{3}{2}}$
in P-wave capture, where $M_\ell$ and $E_\ell$ are the magnetic and
electric $\ell$th--pole operators, respectively. 
The superscripts $LSJ$ refer
to the relative orbital angular momentum $L$ between
the $p$ and $d$ clusters, the channel spin $S$ ($S=1/2$ or $3/2$), and
the total angular momentum $J$ (${\bf J}={\bf L}+{\bf S}$),
respectively.  The $M_2$ transition from the $L,S,J=$1,3/2,5/2 capture 
state has been neglected.   

Improvements in the PHH description of the $pd$ continuum
wave functions, particularly in the short-range part, have
led to significative changes in the values of the P-wave capture RMEs than
previously calculated~\cite{Viv96}.  The S-wave capture
RME values have remained essentially unchanged.  

The $E_1$ RMEs have in fact been calculated in two different
ways: firstly, by direct evaluation of the matrix elements of the
$E_1$ multipole operator,

\begin{equation}
\label{eq:e1me}
E_1^{LSJ} = {\sqrt{2} \over \langle JJ_z, 1\lambda | {1\over2}\sigma_3\rangle }
\langle \Psi_3^{\frac{1}{2}\sigma_3} |E_{1\lambda}| 
{\overline \Psi}_{1+2}^{LSJJ_z}\rangle \ ,
\end{equation}

\begin{equation}
\label{eq:e1}
  E_{1\lambda} = {1\over q}  
  \int d{\bf x}\> {\bf j}({\bf x}) \cdot \nabla \times j_1(qx) 
  {\bf Y}_{1\lambda}^{11}(\hat{\bf x})  \ ,
\end{equation}
where  ${\bf j}({\bf x})$ is the nuclear current density
operator, $j_1(qx)$ is the spherical Bessel function of order one,
${\bf Y}_{1\lambda}^{11}(\hat{\bf x})$ are vector spherical
harmonic functions, and the $pd$ wave function is constructed to satisfy
outgoing-wave boundary conditions as in Eq.~(4.3) of Ref.~\cite{Viv96}.  

Secondly, in the long-wavelength-approximation (LWA)
(certainly justified in the energy range under
consideration here) the $E_1$ operator can be expressed as~\cite{FF82}

\begin{equation}
\label{eq:lw}
  E_{1\lambda} \simeq  E_{1\lambda}({\rm LWA1}) 
                     + E_{1\lambda}({\rm LWA2})
                     + E_{1\lambda}({\rm LWA3})\>\>\> ,
\end{equation}
where

\begin{equation}
\label{eq:lw1} 
  E_{1\lambda}({\rm LWA1}) =
  -{\sqrt{2} \over 3} \left[ H \, , \,  
  \int d{\bf x}\> x\,Y_{1\lambda}(\hat{\bf x})
\>\rho ({\bf x}) \right]  \>\>\> , 
\end{equation}

\begin{equation}
\label{eq:lw2} 
  E_{1\lambda}({\rm LWA2}) = { {\rm i}\, q^2 \over 3 \sqrt{2} }
  \int d{\bf x}\, x \, Y_{1\lambda}(\hat{\bf x})\, {\bf x}
\cdot {\bf j}({\bf x}) \>\>\>,
\end{equation}

\begin{equation}
\label{eq:lw3} 
  E_{1\lambda}({\rm LWA3}) =
  {\sqrt{2}\, q^2 \over 15} \left[ H \, , \,  
  \int d{\bf x}\> x^3 \, Y_{1\lambda}(\hat{\bf x})
\> \rho({\bf x}) \right] \>\>\>.
\end{equation}
Here the continuity equation has been used
to relate $\nabla \cdot {\bf j}({\bf x})$
occurring in $E_{1\lambda}({\rm LWA1})$ and $E_{1\lambda}({\rm LWA3})$ to
the commutator $-{\rm i} [ H \, ,\, \rho({\bf x})]$,
where $\rho({\bf x})$ is the charge density operator.  Evaluating
the RMEs of these operators leads to

\begin{equation}
\label{eq:lwme}
E_1^{LSJ} \simeq E_1^{LSJ}({\rm LWA1})+E_1^{LSJ}({\rm LWA2}) \>\> ,
\end{equation}
with

\begin{equation}
\label{eq:lw1me}
E_1^{LSJ}({\rm LWA1}) =
{\sqrt{2} \over \langle JJ_z, 1\lambda | {1\over2}\sigma_3\rangle }
{\sqrt{2}\, q \over 3} \langle \Psi_3^{\frac{1}{2}\sigma_3} | 
\int d{\bf x}\, x Y_{1\lambda}(\hat{\bf x}) \rho({\bf x}) 
|{\overline \Psi}_{1+2}^{LSJJ_z}\rangle \>\>\> ,
\end{equation}

\begin{equation}
\label{eq:lw2me}
E_1^{LSJ}({\rm LWA2}) = 
{\sqrt{2} \over \langle JJ_z, 1\lambda | {1\over2}\sigma_3\rangle }
{ {\rm i}\, q^2 \over 3 \sqrt{2} } \langle \Psi_3^{\frac{1}{2}\sigma_3} | 
\int d{\bf x}\, x Y_{1\lambda}(\hat{\bf x}) {\bf x} \cdot {\bf j}({\bf x})
|{\overline \Psi}_{1+2}^{LSJJ_z}\rangle \>\>\>,
\end{equation}
where only terms up to order $q^2$
have been retained, and it has been assumed
that the initial and final PHH wave functions are exact eigenfunctions of
the Hamiltonian, so that $[H\, ,\, \rho({\bf x})] \rightarrow
- q \, \rho({\bf x})$ in the matrix element.  The contribution to the
$E_1$ RMEs associated with the LWA3 operator, defined in Eq.~(\ref{eq:lw3}),
is of order $q^3$, proviso the assumption above.

By ignoring two-body contributions to the charge density
operator, we further approximate $\rho({\bf x})
\simeq \rho^{(1)}_{\rm NR}({\bf x}) +\rho^{(1)}_{\rm RC}({\bf x})$,
and write correspondingly

\begin{equation}
\label{eq:lwcb}
  E_1^{LSJ}({\rm LWA1}) \simeq  E_1^{LSJ}({\rm LWAc}) 
                              + E_1^{LSJ}({\rm LWAb}) \>\>\>,
\end{equation}
where 

\begin{equation}
\label{eq:LWAc}
E_1^{LSJ}({\rm LWAc})= 
{\sqrt{2} \over \langle JJ_z, 1\lambda | {1\over2}\sigma_3\rangle }
{\sqrt{2}\, q \over 3} \langle \Psi_3^{\frac{1}{2}\sigma_3} | 
\sum_i \epsilon_i\, r^\prime_{i,\lambda} 
|{\overline \Psi}_{1+2}^{LSJJ_z}\rangle \>\>\> ,
\end{equation}

\begin{equation}
\label{eq:lwb} 
  E_1^{LSJ}({\rm LWAb}) =
   {\sqrt{2} \over \langle JJ_z\, 1\lambda|{1\over2}\sigma_3 \rangle }
  {\sqrt{2}\, q \over 3} \langle \Psi_3^{\frac{1}{2}\sigma_3} | 
 \sum_i - { 2 \mu_i-\epsilon_i \over 4 m^2 } (\bbox{\sigma}_i \times {\bf p}_i)_\lambda
  |{\overline \Psi}_{1+2}^{LSJJ_z}\rangle \>\>\>.
\end{equation}
Hence, up to order $q^2$, the $E_1$ RMEs in LWA are given by 
Eq.~(\ref{eq:lwme}), with $E_1^{LSJ}({\rm LWA1})$
defined in Eqs.~(\ref{eq:lwcb})--(\ref{eq:lwb}).
We re-emphasize that the currents used in the present work
satisfy, by construction, the continuity equation with the AV18 interaction, namely
${\bf q} \cdot {\bf j}({\bf q}) =
\left[ T+v_{18} \, , \, \rho^{(1)}_{\rm NR}({\bf q})\right]$.
Therefore, if the contributions $E^{LSJ}_1({\rm LWAb})$ 
and $E^{LSJ}_1({\rm LWA2})$ (the latter of order $q^2$)
were to be negligible, the degree of agreement between the $E_1({\rm LWAc})$ and
full current results, obtained directly from Eq.~(\ref{eq:e1me}), would simply
reflect the extent to which the present variational wave functions are
truly exact eigenfunctions of the AV18/UIX Hamiltonian.
In Table~\ref{tab:lwa} the results for the $E_1$ RMEs obtained by
direct evaluation of the $E_1$ multipole operator matrix elements
given in Eq.~(\ref{eq:e1me}) are compared with those obtained in LWA.
Note that the $E_1^{LSJ}({\rm LWA2})$ contribution, Eq.~(\ref{eq:lw2}),
has been estimated here by using only the spin-part of the IA current operator. 
Of course, single-nucleon convection as well as
two-body currents provide additional
corrections to ${\bf j}({\bf x})$, but these have been ignored in the
evaluation of Eq.~(\ref{eq:lw2me}).
Finally, note that  the RMEs listed in
Table~\ref{tab:lwa} are related to those defined in
Eqs.~(\ref{eq:e1me}) and (\ref{eq:lw1me})--(\ref{eq:lw2me}) via 
\begin{equation}
     \widetilde X_{\ell C}^{LSJ}= 
      \sqrt{ {v_{\rm rel} \over 2\pi\alpha }{\rm
      exp}(2\pi\alpha/v_{\rm rel})} \, 
     {\sqrt{6\pi} \over q \mu_N }
     \, \sqrt{4\pi} \, X_\ell^{LSJ} \label{newRME2} \>\>,
\end{equation}
where $\mu_N$ is the nuclear magneton and $v_{\rm rel}$ is the $pd$ relative
velocity.  The quantities $\widetilde X_{\ell C}^{LSJ}$ are easily shown to
remain finite in the limit $v_{\rm rel} \rightarrow 0$.

As can be seen by inspection of Table~\ref{tab:lwa}, for the $S=1/2$
states the values of the LWAc and \lq\lq full\rq\rq $E_1$ RMEs are very
close to each other and the remaining small differences between
them are presumably due to the approximate
calculation of the $E_1^{LSJ}({\rm LWA2})$
contribution carried out here (see above).  Therefore, the
present PHH wave functions for these P-wave
channels are a good approximation to the true
eigenfunctions.  This is to be contrasted with the
results reported in Table VI of Ref.~\cite{Viv96}, where
significant differences remained particularly for the
$S=1/2$ $J=1/2$ $E_1$ RME.

The situation is different for the $S=3/2$ states.  Here, as
discussed in Ref.~\cite{Viv96} and below, due to cancellations among
different contributions, the $E_1({\rm LWAc})$ RMEs are rather small
and indeed have similar magnitude as the $E_1({\rm LWAb})$
and  $E_1({\rm LWA2})$ RMEs.
Hence, if the PHH wave functions for these channels also approximate well 
the true eigenfunctions-and indeed there is no reason to 
believe that this is not so-, then the differences between the 
\lq\lq full\rq\rq and LWAc values provide a \lq\lq measure\rq\rq 
of the corrections beyond the standard LWAc. 
Inspection of Table~\ref{tab:lwa} in fact shows that the leading LWAc 
form of the $E_1$ operator is inadequate for these channels. 
This point will be further elaborated below and in the 
next subsection. 

As a final remark, note that the continuity equation
requires the presence of three-body currents
associated with the three-nucleon interaction.  These currents have
been studied in Ref.~\cite{Mar98}, where they were found to give a
very small contribution to the trinucleon magnetic moments and form
factors.  It is, therefore, unlikely that three-body current
contributions influence significantly the
\lq\lq full\rq\rq predictions discussed above,
as it was speculated in Ref.~\cite{Viv96}.

The leading RMEs are the doublet and quartet $M_1$ in S-wave
capture and doublet (namely, $S=1/2$) $E_1$'s with $J=1/2$ and $3/2$
in P-wave capture.  The $M_1$ and doublet $E_1$ strengths are comparable.
The $E_2$ RME is more than an order of magnitude smaller than any
of the two $M_1$'s, the quartet (namely, $S=3/2$) $E_1$'s with $J=1/2$
and $3/2$ are an order of magnitude smaller than the doublet $E_1$'s, and
the $M_2$ strength is negligible.  
In LWAc the $E_1$-multipole operator
is spin-independent, and transitions from the $^3$He ground state
to the $S$=3/2 channel $pd$-states are inhibited, since they must
proceed through the relatively small D-wave components of the
$^3$He wave function.  Hence, the quartet $E_1$ RMEs are
individually small.  Such is not the case for the doublet
$E_1$ RMEs, which result from transitions involving the
$S$=1/2 $pd$ states and the dominant S-wave component
of the $^3$He ground state. 

The effects due to two-body currents and
$\Delta$-degrees of freedom are large on the doublet $M_1$
RME--at $E_p=40$ keV, they increase it in magnitude
by 87\%--and significant
on the quartet $M_1$ RME--at $E_p=40$ keV, they reduce it by 10\%.  The
$E_1^{1\frac{1}{2}\frac{1}{2}}$ and $E_1^{1\frac{1}{2}\frac{3}{2}}$
RMEs are increased by 7\% and 10\%, respectively, by two-body
current contributions (at $E_p=40$ keV); however, these contributions
dominate the quartet $E_1^{1\frac{3}{2}J}$, interfering
destructively with the one-body (IA) results, see Table~\ref{tab:lwa}.

The calculated $S$-factor is compared with
the TUNL data~\cite{Sea96,Mea97} in the lab energy range $E_p$=0--300 keV
in Fig.~\ref{fig:S}.  The present results are in better
agreement with data for $E_p$ between 40 and 80 keV 
than previously reported~\cite{Viv96}.  However, at
higher energies the calculation is slightly above the data, in
particular for the data point at $E_p$=160 keV.  Below 40 keV
theory is again slightly above the data of Ref.~\cite{Sea96}.

The calculated $S$-factor value in S-wave capture at zero energy
is $0.110$ eV-b, in excellent agreement with
the corresponding experimental value
$0.109 \pm 0.010$ eV-b~\cite{Sea96}.  However,
the calculated value in P-wave capture is
$0.109$ eV-b, which substantially overpredicts the experimental value
$0.074 \pm 0.01$ eV-b~\cite{Sea96}.  This overprediction
at zero energy (and below $40$ keV) is due to excess strength
in the calculated $L,S,J=1,1/2,3/2$ $E_1$ RME, 
as discussed below.  

The predicted angular distributions of the differential
cross section $\sigma(\theta)$, vector and tensor analyzing powers
$A_y(\theta)$ and $T_{20}(\theta)$, and photon
linear polarization coefficient $P_\gamma(\theta)$ are compared with
the TUNL data from Ref.~\cite{Sea96} in Fig~\ref{fig:cpt080keV}.
As in Ref.~\cite{Viv96}, we have integrated the theoretical
results, weighted with the energy dependence of the cross section
and target thickness, for the 
purpose of comparing them with experiment~\cite{RSW95}.  
The $\simeq$ 10\% changes in the $E_1$ RMEs, due to the use of more
accurate PHH wave functions, are responsible for the improved description
of the vector analyzing power $A_y$ which was significantly
underpredicted by theory in our earlier work~\cite{Viv96}.  However, the
calculated $\sigma(\theta)$ ($T_{20}(\theta)$) is much 
smaller (larger) than the experimental values at small angles.  The
small angle discrepancy for $T_{20}$ was also present in Ref.~\cite{Viv96},
although it was not as pronounced as found here. 

Recently, the TUNL group has measured additional polarization
observables at $E_p$=40 keV~\cite{Wea99}, specifically
the tensor analyzing powers ${\rm i}\, T_{11}$, $T_{21}$,
$T_{22}$ and the circular polarization asymmetry coefficient
$A_\gamma$.  All these observables are found to be in satisfactory
agreement with the present calculations~\cite{Wea99}. 

The unpolarized cross section $\sigma(\theta)$ and the
tensor analyzing power are given by  
\begin{equation}
  \sigma(\theta)  =\sum_{\ell=0} 
    a_\ell P_\ell(\cos\theta)\ , \qquad
  \sigma(\theta)
   T_{20}(\theta)=\sum_{\ell=0} 
   c_\ell P_\ell(\cos\theta) \>\>\>,
\end{equation}
where the $P_\ell$ are Legendre polynomials, and
the coefficients $a_\ell$ and
$c_\ell$ can be expressed in terms of the various RMEs~\cite{Viv96}.
Hereafter, for ease of presentation, we introduce the notation: 

\begin{eqnarray}
    m_{2J+1}&=&  M_1^{0 J J } \ ,\label{mmpppp1} \\
    p_{2J+1}&=&  E_1^{1 {1\over2} J } \ , \label{mmpppp2} \\
    q_{2J+1}&=&  E_1^{1 {3\over 2} J } \ , \label{mmpppp3}
\end{eqnarray}
the energy dependence of these RMEs being understood.
The leading coefficients in the expansion of $T_{20}$
are $c_0$ and $c_2$, and their expressions are: 

\begin{equation}
  c_0=  \sigma_1 \,
   \left [ 2\, \Re(p_{2} q_{2}^*) - \sqrt{1\over 2} |q_{2}|^2 
  -\sqrt{2\over 5}\Re(p_{4}  q_{4}^*) +\sqrt{8\over 25}  
  |q_{4}|^2 \right ] \ , 
\label{eq:c0} 
\end{equation}
\begin{eqnarray}
  c_2&=& \sigma_1 \, \biggl [ - \Re(m_{2} m_{4}^*) + \sqrt{1\over 8} |m_{4}|^2 
         -\sqrt{1\over 5} \Re(p_{2} q_{4}^*)+ \sqrt{2} \Re(q_{2} p_{4}^*) 
            \nonumber \\
     &&\qquad +\sqrt{1\over 10} \Re(q_{2} q_{4}^*)+
             \sqrt{1\over 10} \Re(p_{4} q_{4}^*)
         +\sqrt{1\over 8} |q_{4}|^2 \biggr ] \ ,
\label{eq:c2} 
\end{eqnarray}
where

\begin{equation}
  \sigma_1= {4\pi \over 3}\;  { \alpha q \over v_{\rm rel} } 
      \ . \label{eq:sigma_1} 
\end{equation}
Note that the expression for $c_0$ reported in Eq.~(B7) of
Ref.~\cite{Viv96} contains several misprints (the different
expression for $\sigma_1$ used in Ref.~\cite{Viv96} accounts for the
fact that there the Legendre coefficients were given in terms
of the RMEs $\widetilde X^{LSJ}_\ell$ defined in Eq.~(6.3) of that work).

The coefficients $c_0$ and $c_2$ are found to be rather sensitive to 
the $q_2$ and $q_4$ RMEs (in particular, $c_0$
vanishes if both $q_2$ and
$q_4$ are set to zero).  As discussed above, these RMEs result
from transitions connecting small components of the wave functions,
and are dominated by many-body current contributions
(see Table~\ref{tab:lwa}).  It is not clear, at this
stage, whether the discrepancies
between the measured and calculated $T_{20}$ (particularly pronounced
at small angles) are due
to deficiencies in the wave functions, 
or rather the interactions generating the wave functions, 
and/or the many-body current models.

The expressions for the leading coefficients $a_0$ and $a_2$
in the Legendre expansion of $\sigma(\theta)$ are

\begin{eqnarray}
  a_0&=&  \sigma_1 \left [ |m_{2}|^2+|m_{4}|^2+
            |p_{2}|^2+ |q_{2}|^2+|p_{4}|^2+|q_{4}|^2
          \right ] \ , \label{eq:a0} \\
  a_2&=& \sigma_1\, \left[ \sqrt{2} \Re(p_{2} p_{4}^*)
                           -{1\over2}  |p_{4}|^2
                     -{1\over \sqrt{5}} \Re(q_{2} p_{4}^*) 
                     +{2\over 5}  |q_{4}|^2
                     \right] \ . \label{eq:a2} 
\end{eqnarray}
For these coefficients the contributions of the $q_2$ and $q_4$ RMEs
are completely 
negligible, and the observed small-angle discrepancy between
theory and experiment is due to excess strength in the
calculated $p_4$ RME, see below. 
In particular, we have verified that in this 
low-energy regime the differences between the 
\lq\lq full\rq\rq results and the results obtained by 
using the LWAc values for the $p_2$, $p_4$, $q_2$ and $q_4$ RMEs 
are small for all the observables considered in Fig.~\ref{fig:cpt080keV}.  
This is not the case for the $pd$ capture at 2 MeV, 
as discussed in the next subsection.

The extensive body of data measured at TUNL has allowed 
the determination of the leading $M_1$ and $E_1$ RMEs (magnitudes
and phases) via fits to the measured observables~\cite{Wea99,RSW95}.  The
results of this fitting procedure~\cite{RSW95} are compared with the
calculated RMEs in Table~\ref{tab:rmew}.  The
RMEs listed in Table~\ref{tab:rmew} have been further multiplied by
the factor
\begin{equation}
    \tilde x_{2J+1}=  \sqrt{  32 \pi \alpha q \mu p \over 2J+1 }  
           x_{2J+1} \ ,
                  \label{eq:Xwel}
\end{equation}
where $\mu$ is the $pd$ reduced mass and $\alpha$ is the fine structure
constant, in order to match the different definitions used in the
present study and by the authors of the fit.  Here, $x_{2J+1}$ stands for
either $m_{2J+1}$, $p_{2J+1}$ or $q_{2J+1}$.  Note that
the phase of each RME is simply related to
the elastic $pd$ phase shift $\delta^{LSJ}$, as discussed in
Refs.~\cite{SK99,watson}.  In particular, at these low energies to
a very good approximation $\delta^{LSJ}\simeq \sigma_L$,
where $\sigma_L$ is the $L$-wave Coulomb phase shift.  For example, the
phase of the calculated doublet $p_2$ RME at $E_p=35$ keV
is found to be $22.648^o$, which is to be compared with an elastic
$pd$ phase shift $\delta^{1{1\over2}{1\over2}}$ of $22.635^o$
(at this energy, $\sigma_1=22.625^o$).  As can be
seen from Table~\ref{tab:rmew}, the most
significant differences between theoretical and experimental RMEs are 
found for $|\tilde p_4|$.  

The experimental value of the quartet to doublet ratio,
$r_{M1}\equiv 2|\tilde m_4|^2/|\tilde m_2|^2$,
for the $M_1$ strength had been
determined to be $0.49\pm 0.04$ in Ref.~\cite{Rea97} from
an analysis of the $T_{20}$ data at $90^\circ$.  The value
obtained from the \lq\lq experimental \rq\rq $M_1$ RMEs listed
in Table~\ref{tab:rmew} is
$0.43\pm 0.05$.  Both of these are in
good agreement with the theoretical prediction of $0.475$. 

It is interesting to analyze the ratio
$r_{E1} \equiv |\tilde p_4|^2/|\tilde p_2|^2$.
Theory gives $r_{E1}\simeq 1$, while from 
the fit it results that $r_{E1}\approx 0.74\pm0.04$.
As discussed above, the calculation of these RMEs is not
influenced by uncertainties in the two-body currents, since their
values are entirely given by the LWAc form of the $E_1$ operator, which
has no spin-dependence.  It is therefore
of interest to examine more closely the origin of the above discrepancy.
If the interactions between the $p$ and $d$ clusters are switched off 
(hence reducing the $pd$ scattering wave function $\Psi_{1+2}^{L=1,SJJ_z}$
in Eq.~(\ref{eq:psia}) to the product of a deuteron wave function
times $Y_1(\hat{\bf r}_{pd}) F_1(q r_{pd})$), the relation $r_{E1}
\simeq 1$ then simply follows from angular momentum algebra, apart
from negligible corrections due to the small P-wave components of the
$^3$He wave function.  Deviations of this ratio from one are therefore
to be ascribed to differences induced by the interactions in the
$L,S,J=1,1/2,1/2$ and $1,1/2,3/2$ wave functions.  To study these
differences, we define the \lq\lq density functions\rq\rq \ 
$\tilde p_{2J+1}(r_{pd})$ with the property 
\begin{equation}
\label{eq:rpd}
\tilde p_{2J+1}=\int_0^\infty dr_{pd}\, \tilde p_{2J+1}(r_{pd}) \>\>\>.
\end{equation}
In Fig.~\ref{fig:prd}, the functions $\Re[\tilde p_2(r_{pd})]$
and $\Re[\tilde p_4(r_{pd})]$ are displayed with and without
including intercluster interactions.  As
expected from the analysis above, 
the two functions $\Re[\tilde p_{2J+1}(r_{pd})]$ are indistinguishable
when these interactions are ignored.  When the latter are included,
the functions $\Re[\tilde p_{2J+1}(r_{pd})]$
are shown by the thick dashed ($J=1/2$) and thick solid ($J=3/2$)
lines.  Intercluster interactions have a significant effect reducing,
however, both integrated values $\tilde p_{2J+1}$ by the
same amount $\simeq 10$\%, with the result that
$\Re[\tilde p_2]$ is still $\simeq \Re[ \tilde  p_4]$. 
 
The AV18/UIX interactions produce essentially
the same asymptotic behavior  
in the $J$=1/2 and $J$=3/2 doublet P-wave scattering states.  This is directly
confirmed by a comparison of the calculated nuclear $pd$ 
elastic phase shifts with the AV18/UIX Hamiltonian model in the
${}^2P_{1/2}$ and ${}^2P_{3/2}$ channels, 
as shown for a few incident proton
energies in Table~\ref{tab:epd}.  The ${}^2P_{1/2}$ and 
${}^2P_{3/2}$ phase
shifts are found to be very  close to each other over the whole energy
range considered in the table.  Note that the AV18/UIX predictions for
these phase shifts at $E_p=3$ MeV are in excellent agreement with the
values extracted from the phase-shift analysis (PSA) performed in
Ref.~\cite{KRTV96}. 

Therefore, the \lq\lq experimental\rq\rq value
$r_{E1}=0.74\pm 0.04$
is at variance with predictions based on
the AV18/UIX Hamiltonian model.  It should be emphasized that
the present study ignores, in the continuum states, the
effects arising from electromagnetic interactions
beyond the static Coulomb interaction between protons.
It is not clear whether the inclusion of these
long-range interactions, in particular their spin-orbit component,
could explain the observed splitting between
the $\tilde p_2$ and $\tilde p_4$
RMEs.  A calculation incorporating them in the continuum
is currently underway, and it
will be the subject of a forthcoming paper.

Finally, the calculated $A_{y}$ and $T_{20}$ analyzing powers 
at $E_{{\rm{c.m.}}}=75$ keV and 100 keV are compared 
with the data of Ref.~\cite{Mea97} in Fig.~\ref{fig:ayt20}. 
Note that the theoretical predictions have 
changed slightly with respect to the earlier results in Ref.~\cite{Mea97}, 
due to the improvement in the description of the $pd$ P-waves discussed
previously. Somewhat better (worse) agreement between 
theory and experiment is now found for $T_{20}$ ($A_y$) than previously
reported~\cite{Mea97}. The large discrepancy at small angles for
$T_{20}$ is present also at these energies.

\subsection{The $pd$ capture at 2 MeV}

Measurements of capture polarization observables
at $E_{\rm c.m.}=2$ MeV have been
reported recently by Smith and Knutson~\cite{SK99}, who also
extracted, by a fitting procedure, values of the contributing RMEs at this
energy.  To reduce the number of parameters in the fit,
they made use of both $pd$ elastic scattering and radiative capture
data.  In fact, as discussed in Ref.~\cite{watson}, the phases of the radiative
capture RMEs are fixed by the elastic $S$-matrix, when no channels other than
elastic or radiative capture ones are open.  Furthermore,
by using the invariance of the nuclear Hamiltonian under parity and
time-reversal transformations, it can be shown that the quantities~\cite{SK99}
\begin{equation}\label{eq:RP}
  \overline{\cal E}^{LSJ}_\ell =  {\rm e}^{-{\rm i} \delta^{LSJ}}
    \sum_{L'S'} U^J_{LS,L'S'} \overline E^{L'S'J}_\ell\ ,
  \qquad
    \overline{\cal M}^{LSJ}_\ell = {\rm e}^{-{\rm i} \delta^{LSJ}}
    \sum_{L'S'} U^J_{LS,L'S'} \overline M^{L'S'J}_\ell\ ,
\end{equation}
are real.  In
the above expression $\delta^{LSJ}$ are the eigenphase shifts and 
$U^J_{LS,L'S'}$ is the mixing matrix, into which the
$S$-matrix describing $pd$ elastic scattering can be decomposed
(both the eigenphases and mixing matrix elements are real
under the deuteron breakup threshold).  The 
RMEs $\overline X^{LSJ}_\ell$ ($X=E$, $M$) used in the expression
above are related to the RMEs $X^{LSJ}_\ell$ defined in the 
present work (see the Appendix) and 
in Ref.~\cite{Viv96}  by 
\begin{equation}
\label{eq:RR}
  \overline X^{LSJ}_\ell = {\rm i}^{L+\ell}\> (-1)^{L+S-J} 
  2\,p\sqrt{q\over v_{\rm rel}} { X^{LSJ}_\ell \over \sqrt{2J+1}}
   \ .
\end{equation}
In Ref.~\cite{SK99}, the real quantities $\overline {\cal E}$
and  $\overline {\cal M}$ have
been used as free parameters and extracted from the data.  The measured
observables, though, were not sufficient to univocally select the values of 
the RMEs and two sets of parameters have been
presented~\cite{SK99}.  In order to compare directly
with the quantities $\overline{\cal E}$ and  $\overline {\cal M}$, we
have transformed our RMEs via Eqs.~(\ref{eq:RP}) and~(\ref{eq:RR}), using the
eigenphase shifts and mixing matrices predicted by the AV18/UIX
model.  The RMEs obtained in this way are reported 
in Table~\ref{tab:rmeP} along with the two sets of \lq\lq experimental \rq\rq
values.  Inspection of the table indicates that
significantly better overall agreement exists between
\lq\lq set 1 \rq\rq and the theoretical RMEs.  
We will only consider this set in
the discussion to follow.  

At this energy, the two doublet $E_1$'s are the dominant RMEs.  They
have similar values, and are in good
agreement with those extracted from the fit.
Note that $\overline {\cal E}_1^{1{1\over2}{1\over2}} \simeq
\overline {\cal E}_1^{1{1\over2}{3\over2}}$ corresponds to 
$| \tilde p_2|\simeq |\tilde p_4|$, since the mixing
induced by the matrix $U$ in Eq.~(\ref{eq:RP}) is negligible for
these RMEs.  Therefore, at this energy the relation $ | \tilde
p_2|^2/|\tilde p_4|^2 \simeq 1$ is well verified also by the
\lq\lq experimental \rq\rq RMEs.

The S-wave $M_1$ RMEs ($\overline {\cal M}_1^{0{1\over2}{1\over2}}$ and 
$\overline {\cal M}_1^{0{3\over2}{3\over2}}$) are quite well reproduced by
theory.  Instead, significant differences between theory and experiment are
found for the D-wave $M_1$ RMEs--those extracted from the fit
are an order of magnitude larger than predicted by theory.  It should
be emphasized, however, that most of the observables show 
sensitivity to these RMEs only in the small ($\theta < 30^\circ$)
and large ($\theta > 150^\circ$) angle regions.  However,
using the experimental values for the D-wave $M_1$ RMEs, rather than
the calculated ones, does not produce
\lq\lq theoretical \rq\rq observables in significantly better agreement with
data. 

The \lq\lq experimental \rq\rq quartet P-wave $E_1$
RMEs ($\overline {\cal  E}_1^{1{3\over2}{1\over2}}$ and  $\overline {\cal
E}_1^{1{3\over2}{3\over2}}$) are overestimated by
theory by almost a factor of $3$.
These RMEs are extremely sensitive to two-body currents,
as can be seen in Table~\ref{tab:lwP}.  The values obtained in the
LWAc, LWAb and LWA2 approximations are also listed in Table~\ref{tab:lwP}.
Similar considerations to those discussed in the previous
subsection apply to these RMEs.  Here we only
point out that the LWAc values of the quartet
$E_1$ RMEs are rather close to corresponding values extracted from the
data (set 1).  This is particularly evident when the comparison is
performed directly in the observables, as can be seen from
Fig.~\ref{fig:cpt2MeV}.  In the figure, the dotted and thin solid curves
are obtained in IA and by using the
full current, respectively, while the thick solid curves are obtained by
retaining the values calculated in the LWAc approximation for
all electric dipole RMEs. 

Note that for the observables
$\sigma$, $A_y$, and ${\rm i} T_{11}$, the contributions
of the quartet $E_1$'s are negligible.  These observables
depend mainly on the doublet $E_1$'s and S-wave $M_1$'s,
and are well predicted by theory.  In contrast,
the observables $T_{20}$ and $T_{21}$ depend
linearly on the quartet $E_1$'s, and are not well reproduced 
by the \lq\lq full\rq\rq theory. 
The differences between the \lq\lq full\rq\rq
and \lq\lq experimental\rq\rq values for these RMEs suggest that the 
present model for the two-body currents may have deficiencies.
Note that the LWAc results 
for the electric dipole transitions 
$\overline {\cal E}_1^{1{3\over 2}{1\over 2}}$ and
$\overline {\cal E}_1^{1{3\over 2}{3\over 2}}$ 
happen to be close to the \lq\lq experimental\rq\rq values, as shown 
in Table~\ref{tab:lwP}, and hence good agreement is obtained 
between the experimental and LWAc-calculated $T_{20}$ and $T_{21}$ 
observables. It is, however, important to emphasize 
that such an agreement is to be considered purely accidental, since the 
next to leading order contributions LWAb and LWA2 are 
comparable to the leading order LWAc results.

Some of the differences found between the theoretical
and experimental RMEs could be due to the different elastic eigenphase
shift and mixing angle parameters, used in Eq.~(\ref{eq:RP}) to 
define the real RMEs $\overline {\cal E}$ and $\overline {\cal M}$.
To clarify this point, the elastic eigenphase shifts obtained with the
AV18/UIX model are compared with those used
by the authors of Ref.~\cite{SK99} in Table~\ref{tab:eps}.  There is
good overall agreement between the
two sets of parameters.   We note that the
small differences in the ${}^4P$ eigenphase shifts are
responsible for the large underprediction of the $A_y$
and ${\rm i} T_{11}$ observables in elastic $pd$ scattering.

As discussed in the previous section, 
the LWAc and \lq\lq full\rq\rq 
estimates of the quartet $E_1$ RMEs are rather different (see
Table~\ref{tab:lwa}). However, when the LWAc values of these RMEs are used
in evaluating the various observables, the resulting changes are found to
be rather small, even for $T_{20}$ and $T_{21}$. In this energy regime, 
the observables are in fact dominated by the doublet $E_1$ and the S-wave $M_1$
RMEs, and the contributions of the quartet $E_1$ RMEs are found 
to be negligible in all cases. At higher energies, though, the quartet $E_1$
becomes of the same order of magnitude of the S-wave $M_1$ RMEs, 
and therefore have large effects on some of the observables. 
Clearly, the significant differences 
between the \lq\lq full\rq\rq and LWAc predictions for the quartet 
$E_1$ RMEs have a large impact, 
particularly for the tensor observables.

Finally, in Fig.~\ref{fig:S2} the \lq\lq full\rq\rq results for the $S$-factor
are compared to data in the c.m. energy range 0--2 MeV
stored at the web site {\tt http://pntpm.ulb.ac.be/nacre.htm}.

\subsection{Contributions to the Gerasimov-Drell-Hearn integral} 

The Gerasimov-Drell-Hearn (GDH) sum
rule connects the helicity structure of the photo-absorption
cross section to the anomalous magnetic moment of the nuclear target, and
is derived using Lorentz and gauge invariance,
crossing symmetry, causality and unitarity of the forward Compton
scattering amplitude\cite{Ger66,Dre66}.  For the case of $^3$He, it is given by

\begin{equation}
  I_{\rm ^3He}\equiv\int_{\omega_{\rm th}}^\infty d\omega \frac
  { \sigma^\gamma_P(\omega)
  -\sigma^\gamma_A(\omega) }{\omega} = 2 \pi^2 \alpha
  \left(\frac{\kappa_{\,^{3}\!{\rm He}}}{m_{\,^{3}\!{\rm He}}}\right)^2 \>\>,
  \label{gdh}
\end{equation}
where $\sigma^\gamma_{P/A}(\omega)$ are the photon absorption cross
sections in which the photon helicity and $^3$He spin are either
parallel ($P$) or antiparallel ($A$),
$m_{\,^{3}\!{\rm He}}$ and $\kappa_{\,^{3}\!{\rm He}}$ are
the $^3$He mass and anomalous magnetic moment, and $\omega_{\rm th}$
is the threshold energy.  As discussed in the Appendix, 
it is related to the inclusive
$R_{T^\prime}$ response, measured in polarized electron scattering from
a polarized spin 1/2 target. 

In $^3$He, the GDH integral is $I_{^3{\rm He}}= 498\ \mu$b, using
the experimental value $\kappa_{^3{\rm He}}= -8.366$ for its anomalous
magnetic moment.  The photodisintegration
threshold in $^3$He is $\omega_{\rm th}$=5.495 MeV, corresponding to
$pd$ breakup.  It is useful to divide
the integral into the part up to pion production threshold, and the
part above this 
threshold.  For the part above pion threshold,
the $^3$He nucleus should have roughly
the same strength as the neutron, i.e. $\simeq 230$ $\mu$b.  Such an
expectation is based on the fact that the $^3$He
ground state consists predominantly of
a spherically symmetric S-wave component, in which
the proton spin projections are opposite and
the net polarization is therefore
due entirely to the neutron.  Ignoring corrections to this naive estimate,
it is expected that the $\omega$-region from the photodisintegration
threshold up to the pion threshold should contribute
about $266\ \mu$b to the $^3$He GDH integral.  A realistic
description of the $pd$ and $ppn$ continuum states for energies above
the three-body breakup threshold in terms of PHH wave functions, while
certainly possible, is not yet presently available.  A calculation of the 
contribution to the GDH integral from the $ppn$
threshold up to the pion threshold is therefore not possible.  In the present
work, however, we study this contribution in the 2 MeV window where
only two-body photodisintegration channels are energetically allowed.

In fact, the TUNL group has recently made the first experimental
determination of the contribution to the GDH integral of $^3$He 
from the energy region up to 53 keV above the $pd$ threshold~\cite{Wea99}. 
In this region, there are only six dominant RMEs corresponding
to electric and magnetic dipole transitions, as already noted before.
Thus, ignoring the contributions from higher order multipoles, we find
that the cross section difference in Eq.~(\ref{gdh}) is simply given by

\begin{eqnarray}
\label{eq:dsi}
  \Delta\sigma&\equiv& \sigma_P-\sigma_A \nonumber \\
&=& {16\pi^2  \alpha \mu p \over \omega}
    \biggl[ -|m_2|^2+{|m_4|^2\over 2}
-|p_2|^2+{|p_4|^2\over 2} -|q_2|^2+{|q_4|^2\over 2}
    \biggr]\ ,
\end{eqnarray}
in the notation of Eqs.~(\ref{mmpppp1})--(\ref{mmpppp3}).  In
particular, for the purpose
of comparing with the discussion of subsection A, note the factor
$1/\sqrt{2J+1}$ difference between the $\tilde x_{2J+1}$ and $x_{2J+1}$
in Eq.~(\ref{eq:Xwel}).

As discussed in subsection A, the  TUNL group has determined the relevant
RMEs from an analysis  of the polarized capture data, the details of which are
reported in Ref.~\cite{Wea99}.  Because
of time reversal invariance, the RMEs for
the capture reaction are related to those for the
photo-absorption reaction by phase factors, which
are irrelevant for the  $\Delta \sigma$ defined above. 

It is convenient to define
\begin{equation}
\label{eq:I1}
 I(\overline{\omega})= \int_{\omega_{\rm th}}^{\overline{\omega}}
 d\omega {\sigma_P(\omega)-\sigma_A(\omega) \over \omega} \ ,
\end{equation}
with, obviously, $I(\overline{\omega} \rightarrow \infty)=I_{^3{\rm
He}}$.  The experimental values obtained by integrating up to
$\overline{\omega}_1= 5.522$ MeV and $\overline{\omega}_2=5.548$ MeV
are presented in Table~\ref{tab:res}, where they are compared to
predictions obtained by including one-body only and both one-
and two-body currents (columns labelled IA and FULL).
Note that these values represent small negative contributions to
the total strength expected below pion threshold.  Table~\ref{tab:res}
also lists the individual contributions to $I(\overline{\omega}_1)$
from the $-|m_2|^2+|m_4|^2/2$, $-|p_2|^2+|p_4|^2/2$, $-|q_2|^2+|q_4|^2/2$
RME combinations (rows labeled $M_1$, $E_1$ $S$=1/2, and $E_1$ $S$=3/2,
respectively).

The total contributions including the $M_1$, $E_1$ $S$=1/2 and
$E_1$ $S$=3/2 strengths are found to be in IA an order
of magnitude smaller (in absolute value) than data.  This is because in IA
$|m_2|^2 \simeq 0.5 \times |m_4|^2$, $|p_2|^2 \simeq 0.5 \times |p_4|^2$
and the quartet ($S$=3/2) $E_1$ strength is very small.  
 
The ratio $\simeq 0.5$ for the doublet to quartet $M_1$ strength
obtained in IA is consistent with predictions
for $pd$ capture at zero relative energy obtained
with the Faddeev method using a variety of realistic Hamiltonians~\cite{Fea91}
(the ratio is found to have only a weak energy dependence).
When two-body currents are included, the doublet $M_1$ strength
$|m_2|^2$ becomes roughly twice as large as the quartet $M_1$ strength
$|m_4|^2$, a result also consistent with the earlier calculations~\cite{Fea91}.
This makes the overall $M_1$ contribution to $I(\overline{\omega})$
negative and relatively large (in absolute value).  

However, the nearly exact cancellation between the doublet $E_1$ strengths
$|p_2|^2$ and $|p_4|^2/2$ (or $|\tilde p_2|^2$ and $|\tilde p_4|^2$)
discussed earlier, is not significantly influenced
by the inclusion of two-body currents.  The quartet $E_1$
RMEs remain negligible.  Thus the total contribution
to $I(\overline{\omega})$ is mostly due to $M_1$ strength.
While the results in the \lq\lq full\rq\rq calculation are
in much better agreement with data than those in IA, a factor
of two discrepancy persists between theory and experiment.
This discrepancy is mostly due to the difference between
the calculated and measured $S$=1/2 $E_1$ strength, as
the second row in Table~\ref{tab:res} makes clear.
It should also be noted that the measured $I(\overline{\omega}_2=5.548
{\rm MeV})=-1.120 \pm 0.218$ nb is a very tiny piece of the expected
contribution to the GDH integral of $^3$He below pion threshold, i.e.
$266\ \mu$b, and that it has the opposite sign.  

The result for $I(\overline{\omega}_3)$ with $\overline{\omega}_3$=7.7 MeV
corresponding to the threshold for complete breakup into $p$$p$$n$
is 1.1 $\mu$b.  We find that the cross section difference $\Delta \sigma$ changes sign
in the \lq\lq full\rq\rq calculation at about $\omega$=6 MeV.
In the range of energy $6$ MeV
$\le \omega \le 7.7$ MeV the polarized cross section
is dominated by the $E^{1{1\over2}J}_1$
RMEs.  Indeed, the $|p_2|^2$ and $|p_4|^2$ strengths are three orders
of magnitude larger than the strength from any of the other
contributing RMEs.  However, the large cancellation between $-|p_2|^2$
and $|p_4|^2/2$ persists also in this $\omega$-region, although to a
lesser extent than found above: 85\% of $\Delta \sigma$ is due to these
terms, while the remaining 15\% comes from the other RMEs.  Note that
in this region, RMEs other than those included near the two-body
breakup threshold need to be considered.  For example, the electric
quadrupole $E_2^{2{1\over2}{5\over2}}$ at $\omega=7.5$ MeV ($E_p=3$
MeV) is found to give a 4\% contribution to $\Delta\sigma$.

While the results reported above indicate that an
extremely tiny piece of the total sum-rule strength is located in
the threshold region, we find, by direct comparison with
experiment, that this integral observable is very sensitive
to the effects of two-body currents.  The inclusion of these
currents reduces the discrepancy
between theory and experiment from a factor of ten to a factor of two.
Further studies are needed to understand the physical
origin of the difference in the leading P-wave $E_1$ RMEs
responsible for the remaining discrepancy.  
\subsection{Threshold electrodisintegration of $^3$He}

The most recent and systematic experimental study of the
threshold electrodisintegration of $^3$He and $^3$H we are
aware of was carried out by Retzlaff {\it et al.}~\cite{Rea94}
at the MIT/Bates Linear Accelerator Center.  The longitudinal
and transverse response functions $R_L$ and $R_T$ were obtained
using Rosenbluth separations for three-momentum transfers in the
range 0.88--2.87 fm$^{-1}$ and excitation energies from two-body
thresholds up to 18 MeV.  The $^3$H($e,e^\prime$) data are the only
measurements at these energy and momentum transfers.  Inclusive
$^3$He electron scattering data from earlier experiments~\cite{elt}
are in agreement with the Retzlaff {\it et al.} measurements, after
scaling for the slightly different kinematics.

The $R_L$ and $R_T$ $^3$He data at momentum transfer values
$q$=0.88, 1.64 and 2.47 fm$^{-1}$ are compared in Fig.~\ref{fig:res3he}
with calculations using PHH wave functions obtained for the AV18/UIX
Hamiltonian model, and one-body only (dashed lines) or both
one- and two-body (solid lines) charge and current operators. 
Note that the contributions associated with the $L$=0--5 $pd$ scattering
states are retained in the calculation, which is then fully converged.
No calculations of the $^3$H response functions have been
carried out at this time.  There is satisfactory agreement between
theory and experiment for all cases, but for the longitudinal response
at $q$=2.47 fm$^{-1}$.  The two-body components of the electroexcitation
operator play an important role, particularly for the transverse response
at the highest $q$-values.  The relative sign between the one-
and two-body contributions is consistent with that expected
from elastic form factor studies of $^3$He.  There
it is found~\cite{Mar98} that two-body current (charge)
operators increase (decrease) the IA predictions
for the magnetic (charge) form factor at $q \le 3$ fm$^{-1}$.

A feature of the longitudinal $^3$He data is the presence of an
enhancement near threshold.  Such an enhancement is particularly
pronounced at low $q$.  It is not observed in the $^3$H longitudinal
response~\cite{Rea94}.  At the momentum transfers
under consideration here, the longitudinal
strength is almost entirely due to a $C_0$ transition involving the dominant
S-wave component of $^3$He and the doublet S-wave $pd$ scattering
state.  It has been 
shown by Heimback {\it et al.}~\cite{Hei77} that the enhancement results from
the constructive interference of the amplitudes in the two-body breakup of
$^3$He that correspond to the virtual photon coupling directly to a proton
or to a correlated proton-neutron pair.  In $^3$H, the virtual photon
for the two-body breakup channel can couple only to the correlated pair, since
coupling directly to a proton leaves an unbound neutron pair and thus a three-body
final state.  The present calculations correctly account for this threshold
enhancement in the longitudinal response of $^3$He.

Previous calculations, reported on in Ref.~\cite{Rea94}, used either the
Faddeev equations~\cite{MT92} or the
orthogonal-correlated-state (OCS) method~\cite{SP87} to describe
the bound and scattering wave functions.  Both calculations did not include
two-body charge and current operators.  However, the former used the central
Malfliet-Tjon interaction, while the latter was based on a realistic
Hamiltonian including the older Argonne $v_{14}$ two-nucleon~\cite{AV14}
and Urbana-VII three-nucleon~\cite{Sch86} interactions.  The longitudinal
and transverse response functions are surprisingly well  
predicted in the threshold region by the Faddeev calculation, but are
both underestimated, particularly the
transverse response, in the OCS calculation.
The latter also fails to reproduce the observed 
enhancement in longitudinal strength between the two- and three-body
breakup thresholds, probably because of the
approximate treatment of final-state-interaction
effects between the proton and deuteron clusters.  However, in
view of the importance of two-body currents (see Fig.~\ref{fig:res3he}),
the agreement between the data and Faddeev results
is presumably accidental. 

Finally, in Figs.~\ref{fig:rl-rlt} and~\ref{fig:rt-rtt} we show
the $R_L$,  $R_{LT^\prime}$, $R_T$ and $R_{T^\prime}$ response
functions at a fixed excitation energy of 1 MeV above the $pd$
threshold in the three-momentum transfer range 0--5 fm$^{-1}$.
In $R_L$ and $R_{LT^\prime}$ the S-wave $pd$ continuum states
give the dominant contribution, while in $R_T$ and $R_{T^\prime}$
both S- and P-wave states give equally important
contributions over the whole $q$ range.  All response functions
are substantially affected by two-body currents,
however, the sensitivity to these is particularly pronounced
for $R_{LT^\prime}$ and $R_{T^\prime}$. 

\begin{figure}[p]
\caption{The inclusive cross section, and the $A_{LT^\prime}$ and
$A_{T^\prime}$ asymmetries,
obtained with the AV18/UIX Hamiltonian model and one-body only
(dashed lines) or both one- and two-body (solid lines) charge and current
operators, are displayed for $^3$He at a fixed excitation
energy of 1 MeV for three-momentum transfers in the range 0--5 fm$^{-1}$.
The results in PWIA (dotted lines) are also shown.  
The incident electron energy is 4 GeV, and the electron scattering angle is 
in the range 0--14$^\circ$.}
\label{fig:see}
\end{figure}
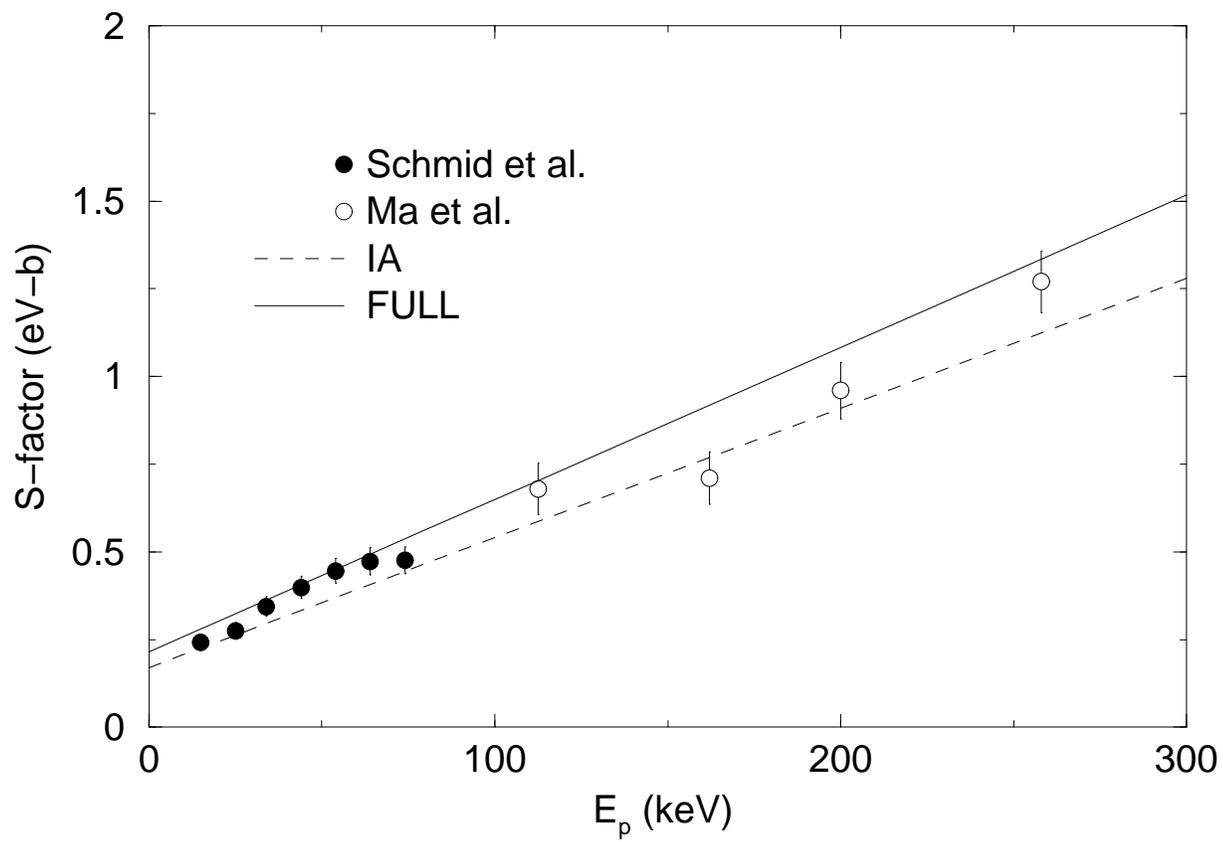
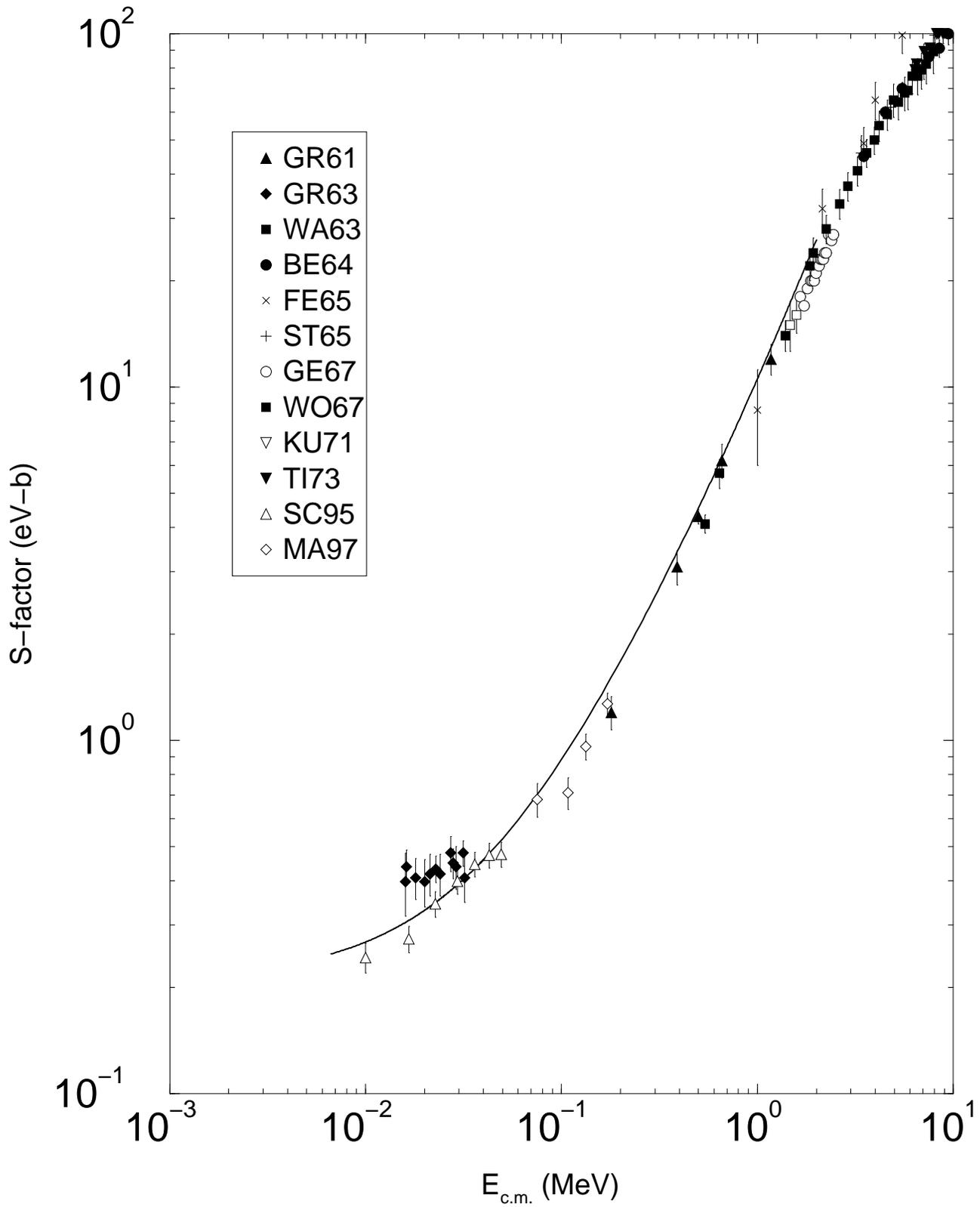
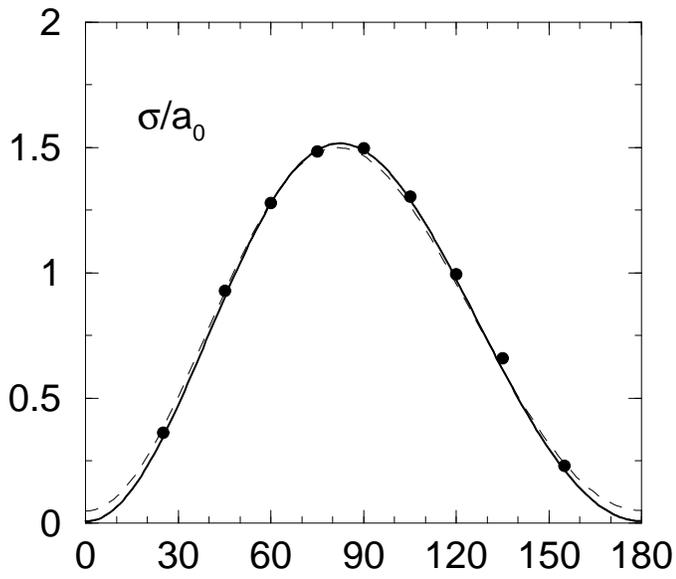
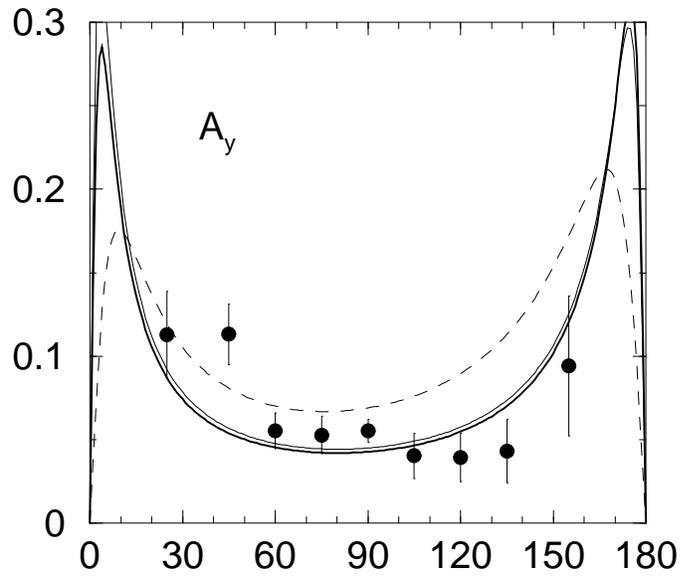
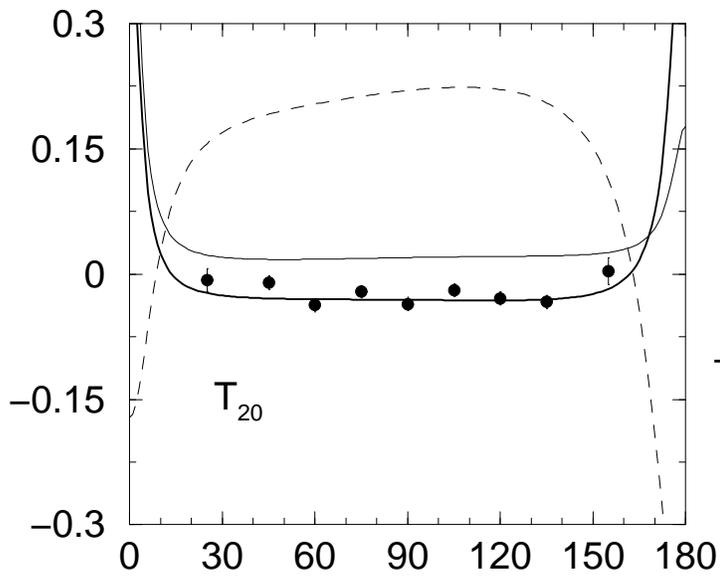
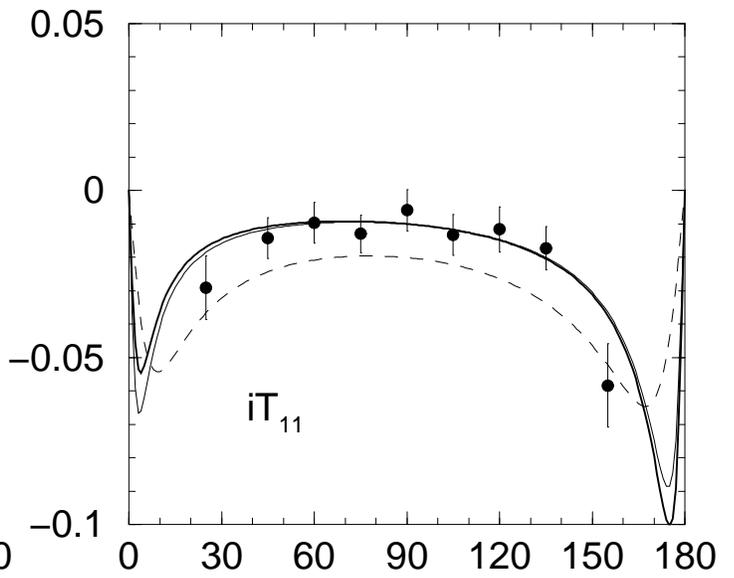
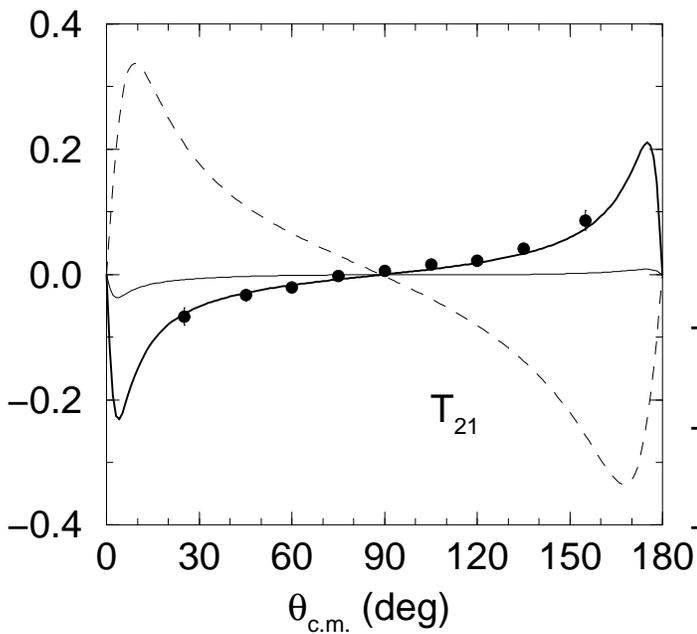
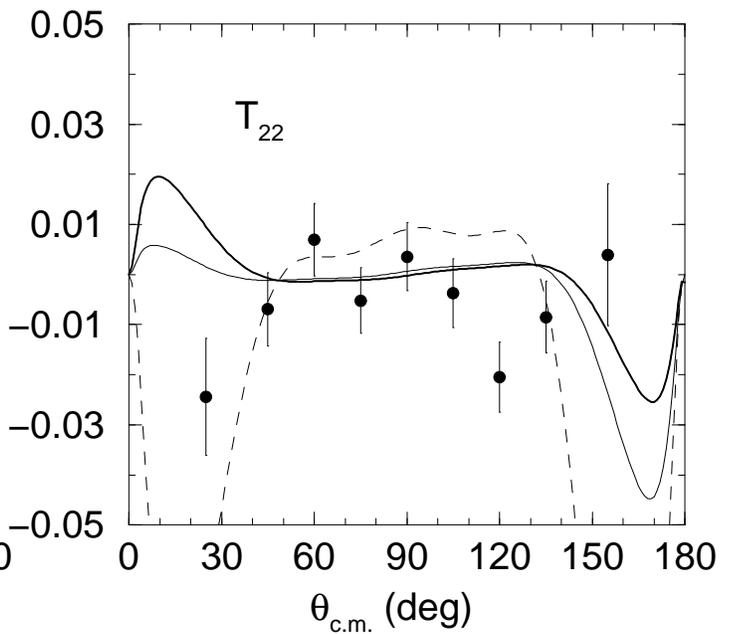
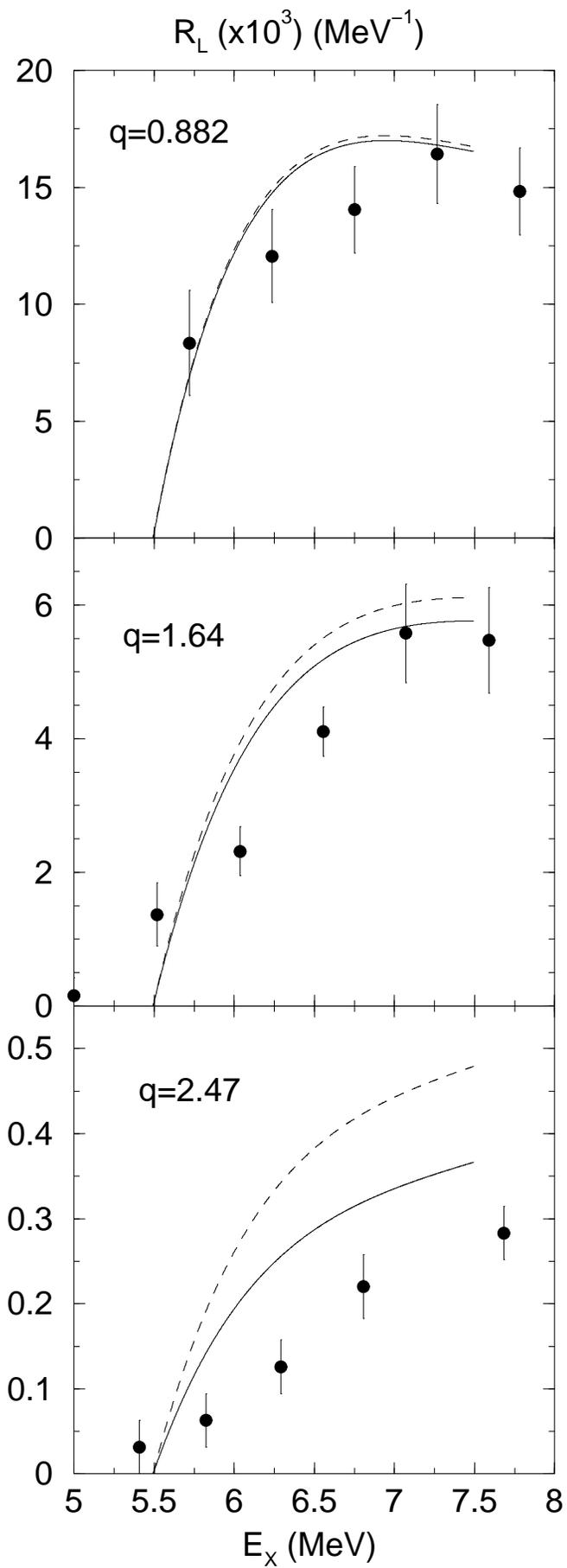
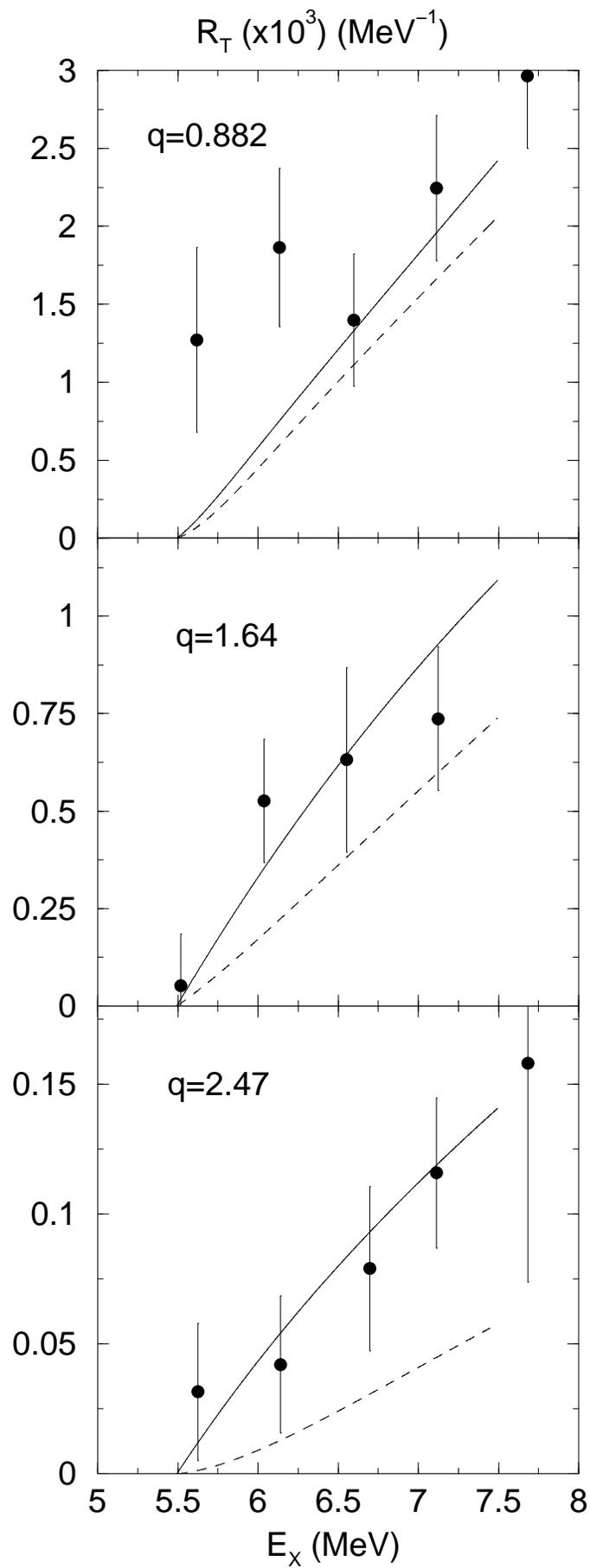
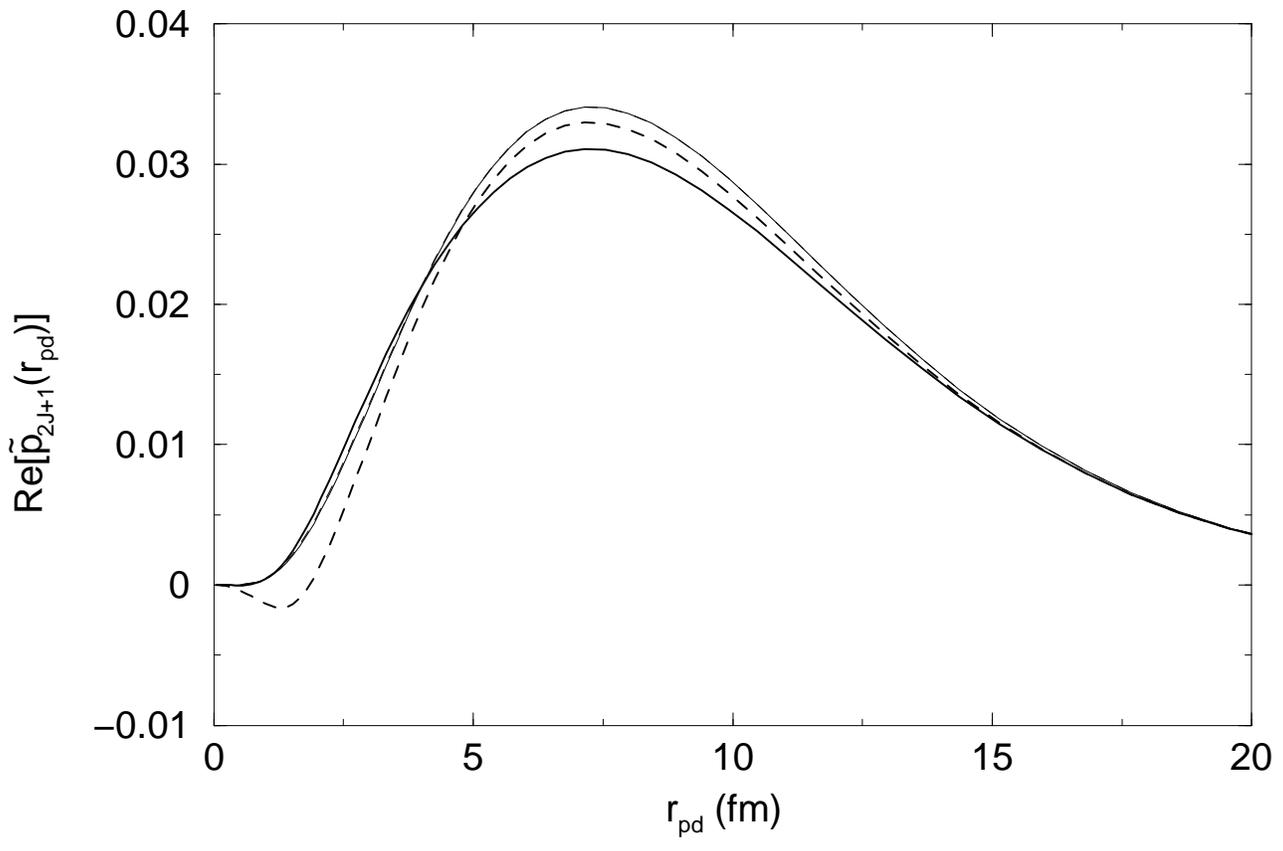
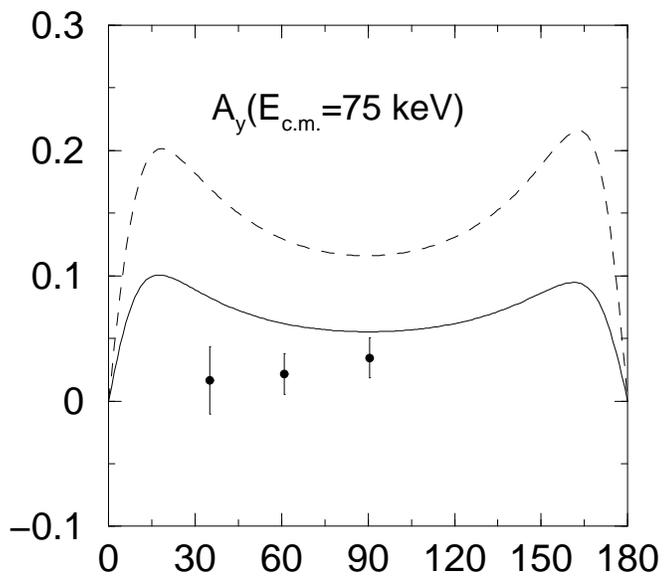
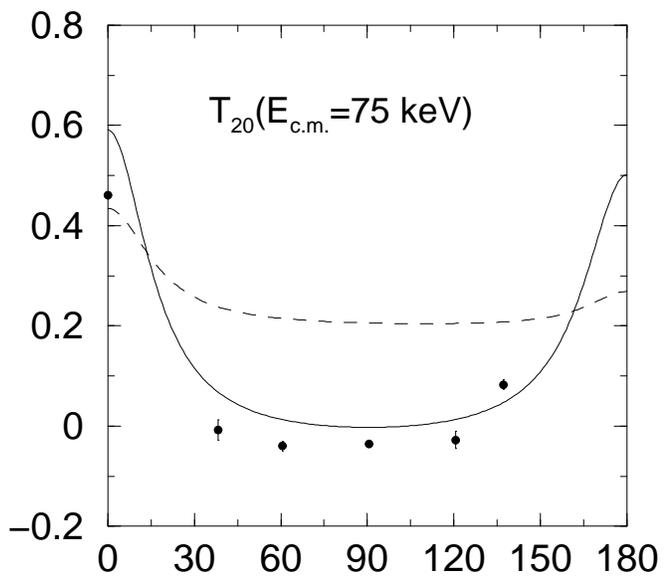
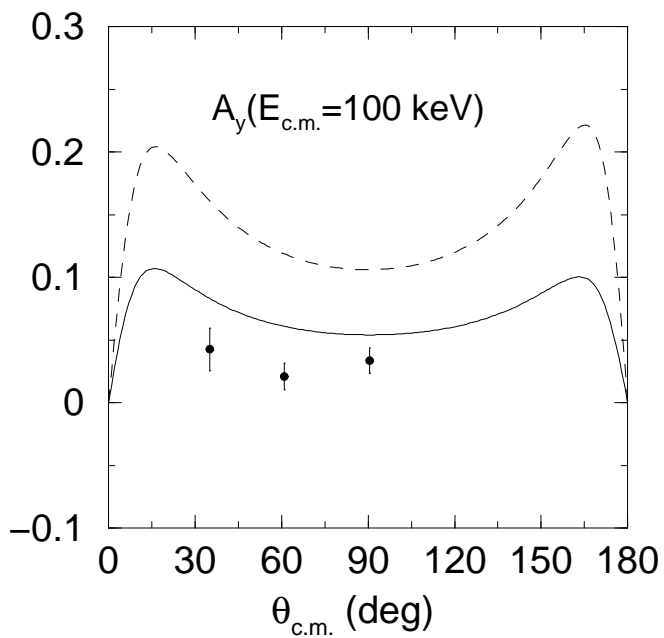
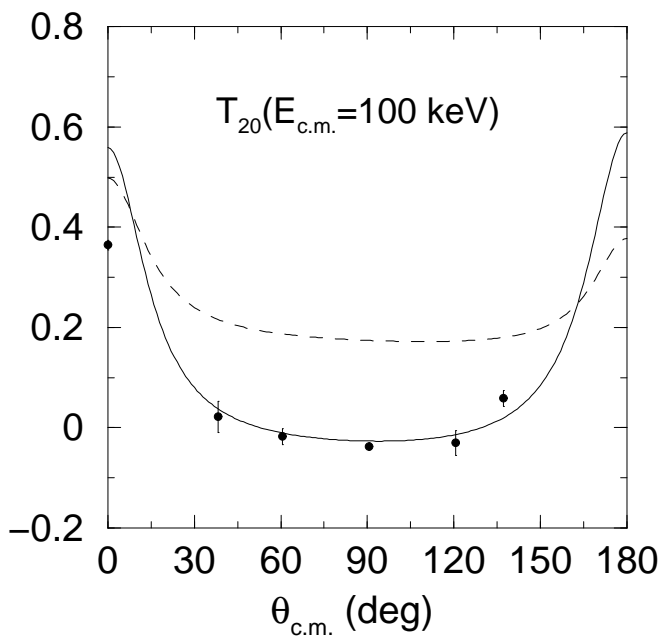
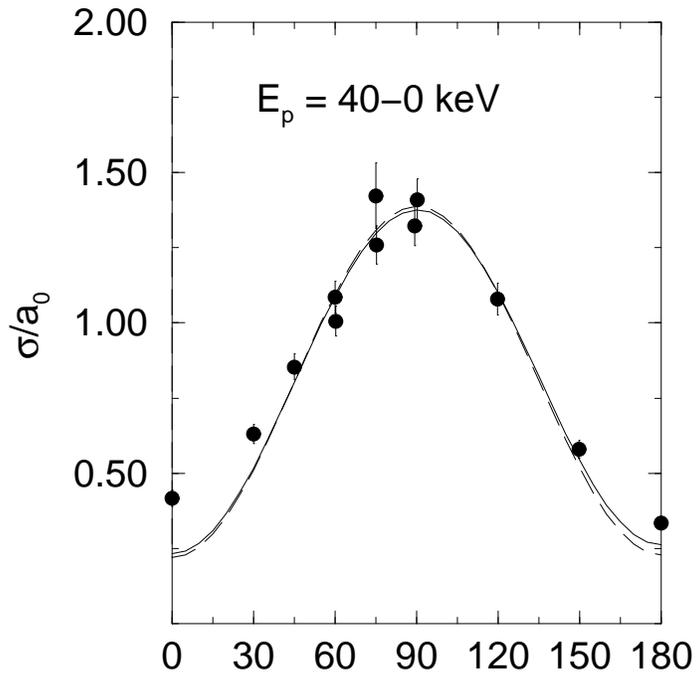
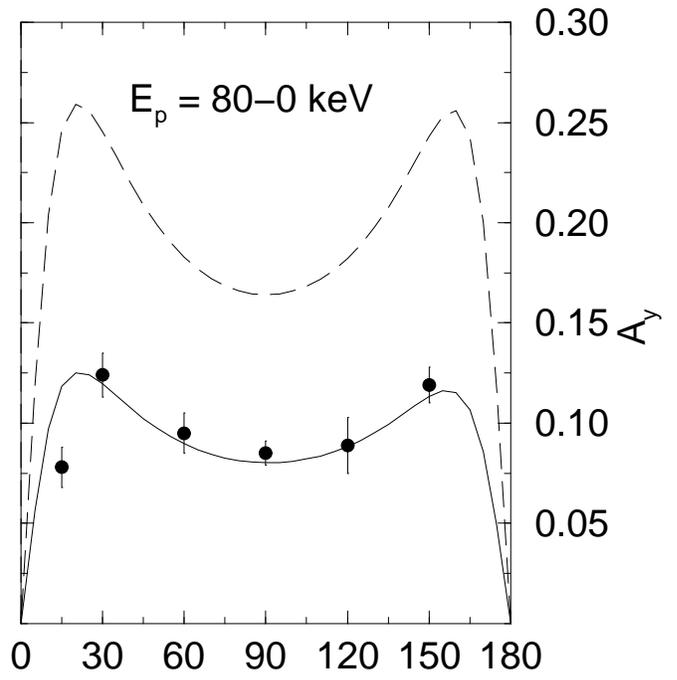
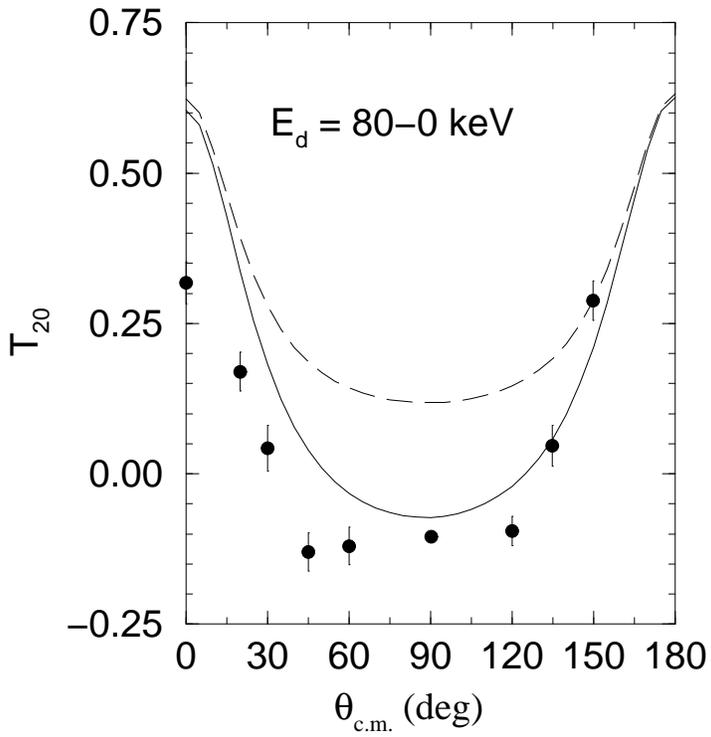
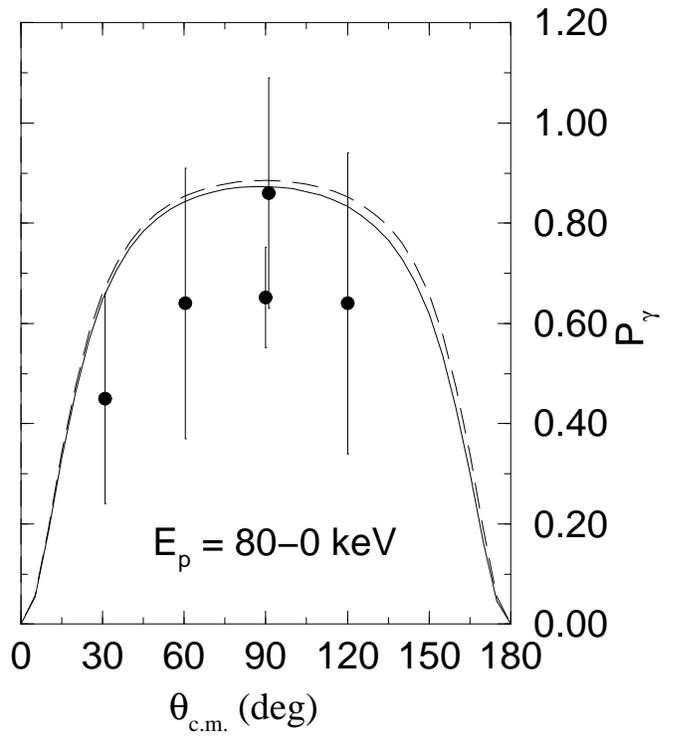
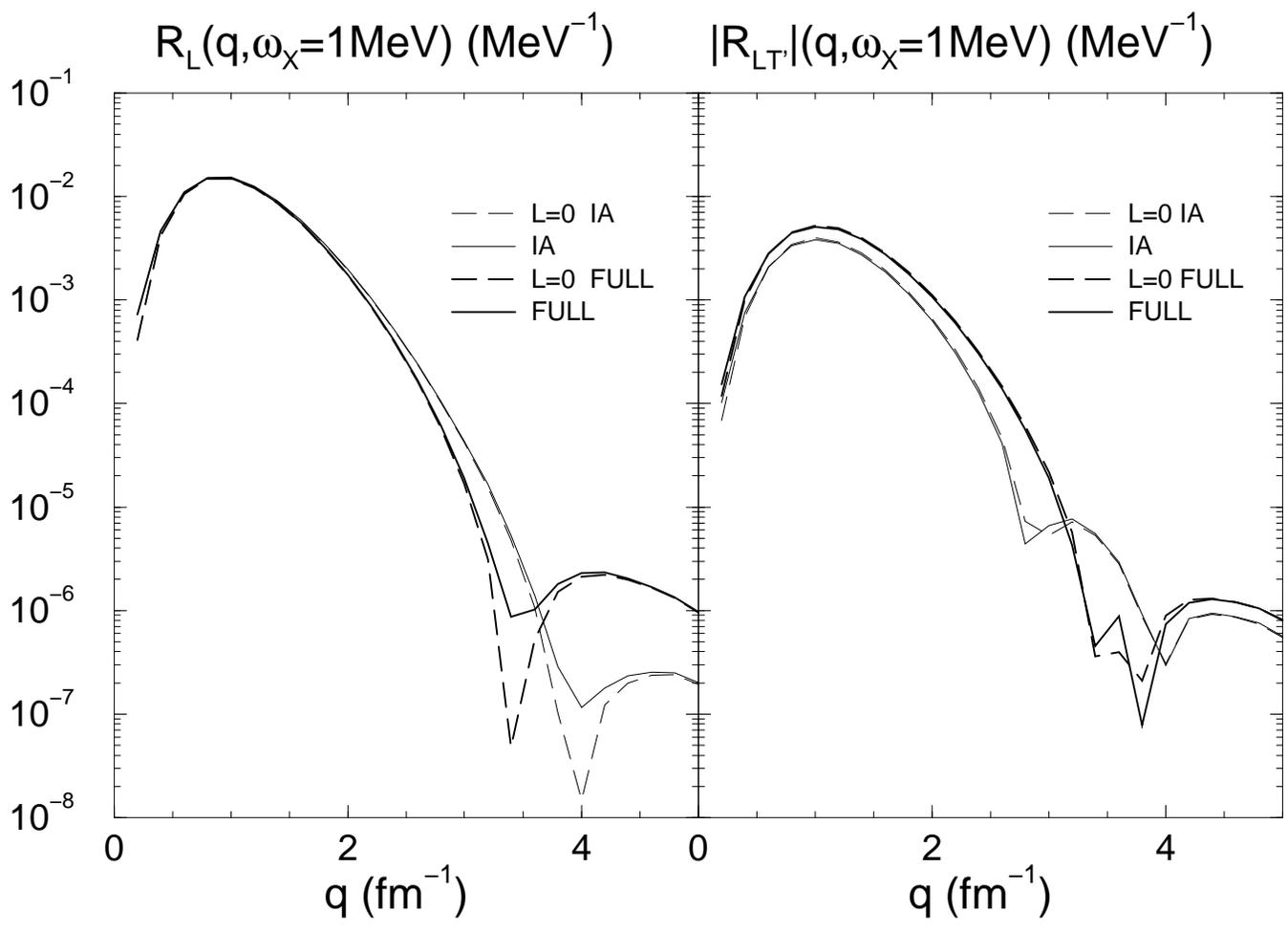
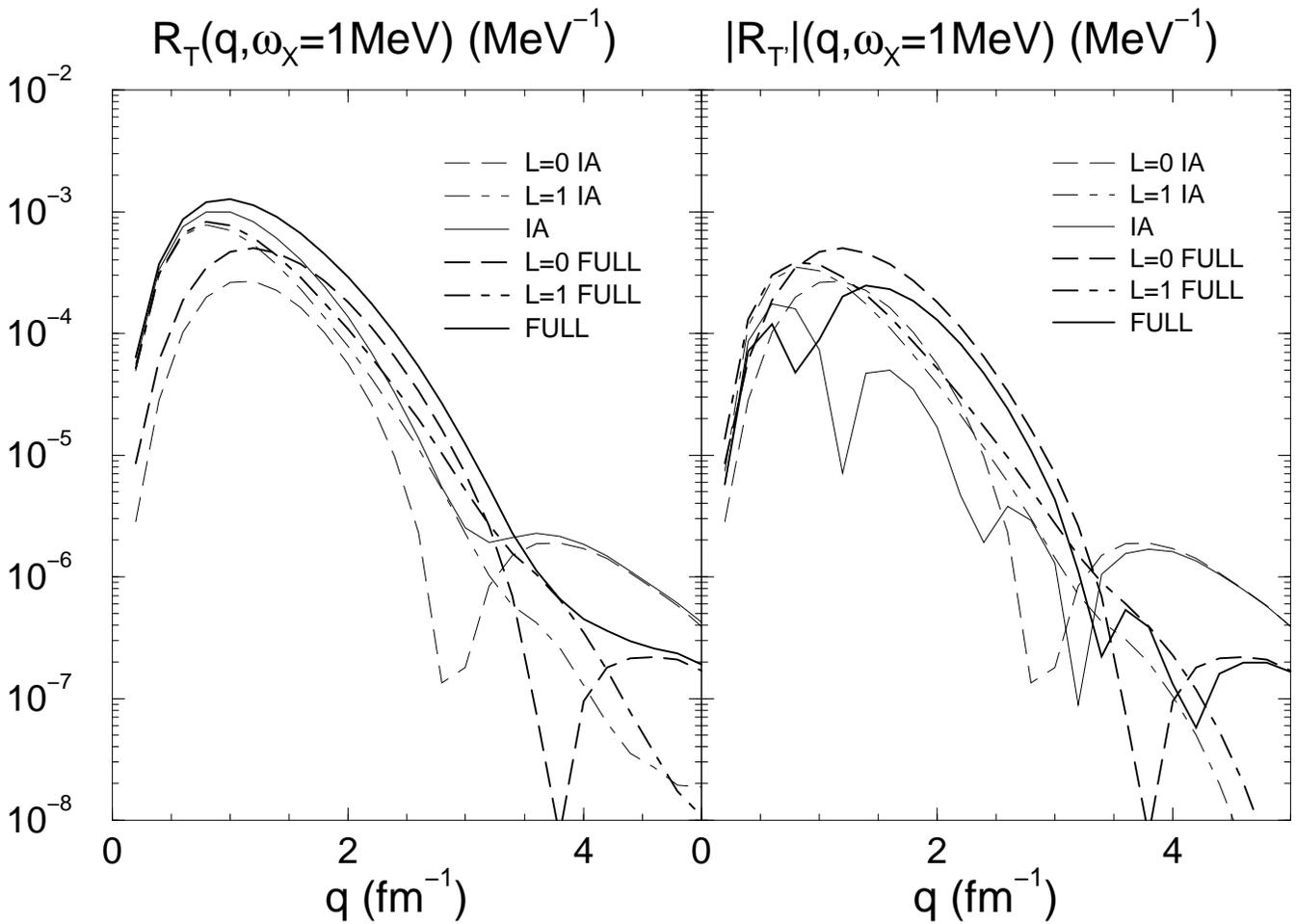
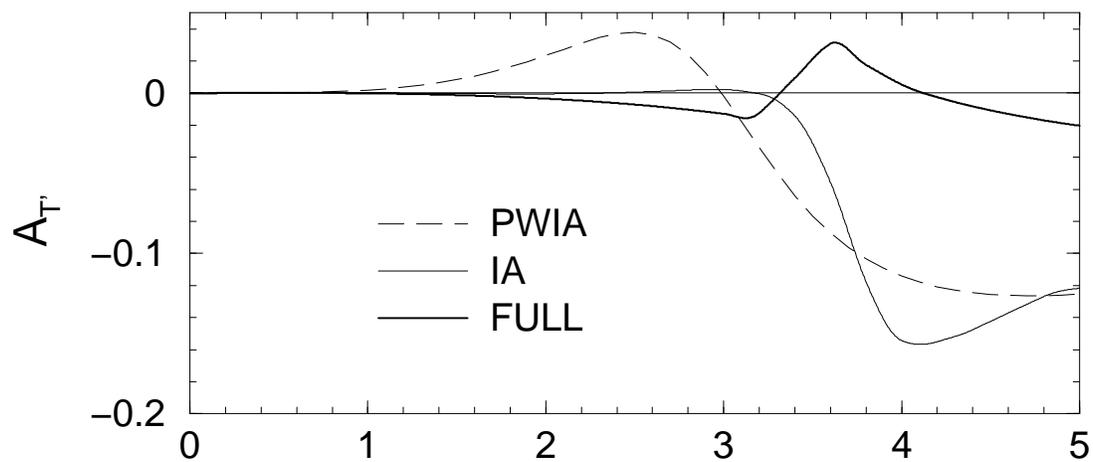
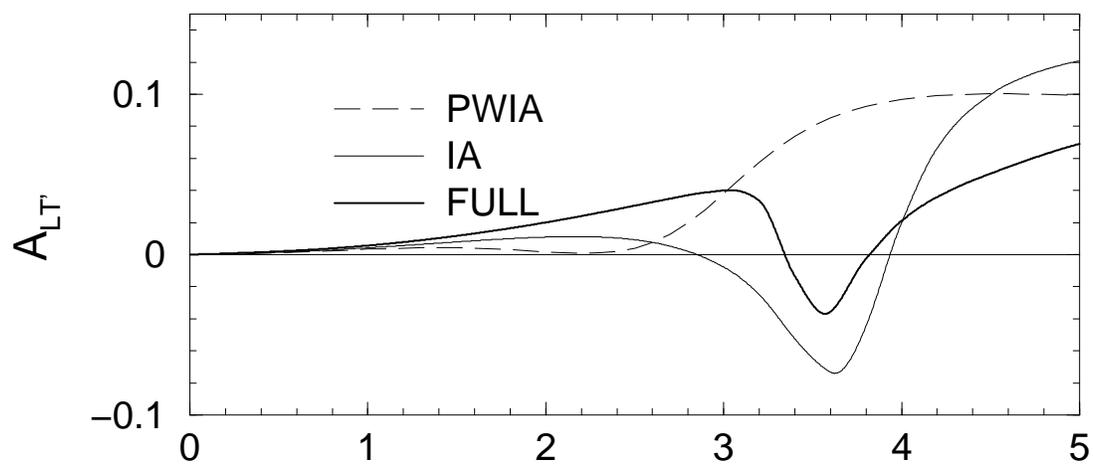
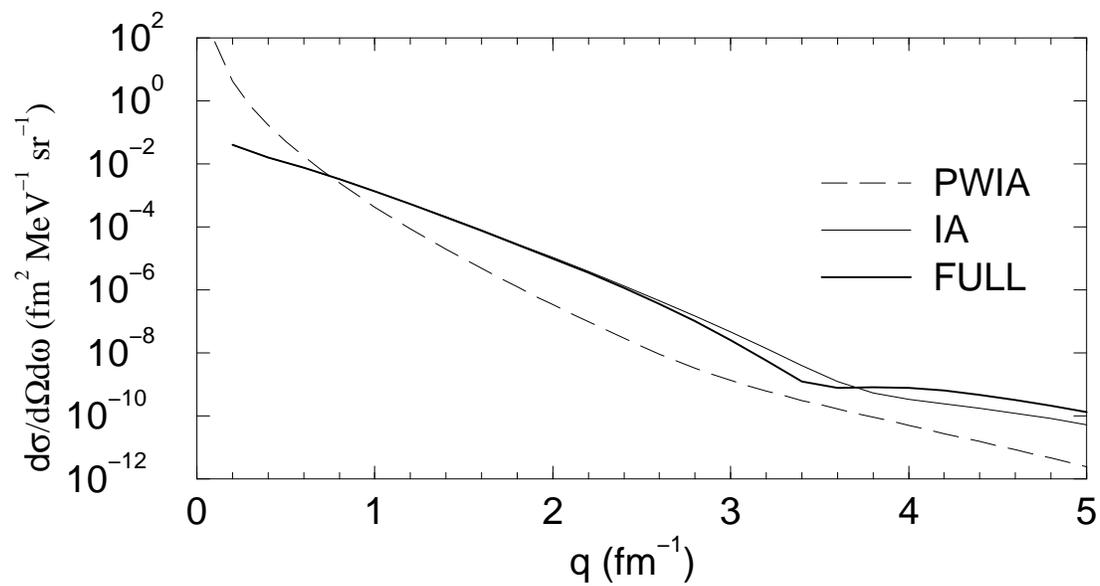

In Fig.~\ref{fig:see} we show the unpolarized cross section, and
the $A_{LT^\prime}$ and $A_{T^\prime}$ asymmetries in the threshold
region at an incident electron energy of 4 GeV.
The asymmetries are relatively large at high $q$, and
particularly sensitive to two-body currents.  Note that $A_{LT^\prime}$
has been found to be little influenced by uncertainties in the
electric form factor of the neutron (the Galster parametrization is used
in Figs.~\ref{fig:res3he}--\ref{fig:see}), except at the highest $q$-values.
The cross section for the chosen kinematics (incident electron 
energy of 4 GeV, fixed $pd$ excitation energy of 1 MeV, and 
$0^\circ\!<\!\theta_{e}\!<\!14^\circ$) is dominated by the longitudinal 
response function. 
Note that in Fig.~\ref{fig:see} we also show the 
plane-wave-impulse-approximation (PWIA) results. These have been calculated 
by approximating the wave function 
\begin{equation}
\Psi_{1+2}^{LSJJ_z}({\rm{PWIA}}) = \sum_{{\rm cyclic}\> ijk}  
   \Big[ \lbrack s_i \otimes \phi_d({\bf x}_i)
   \rbrack_{S} \otimes Y_{L}({\hat {\bf y}}_i) \Big]_{JJ_z}
   {F_{L}(pr_{pd}) \over pr_{pd}} \ .
\label{eq:psipwia}
\end{equation} 
The lack of orthogonality between the $\Psi_{1+2}^{LSJJ_z}({\rm{PWIA}})$ 
with $L=0$ $S=1/2$ 
and the $^{3}$He ground-state wave functions is responsible for the excess 
cross section at low $q$ obtained in PWIA with respect to the 
\lq\lq full\rq\rq calculation.
\section{Conclusions}

We have reported calculations of
$pd$ radiative capture observables at energies
below the three-body breakup threshold, and of
longitudinal, transverse and interference response
functions measured in polarized electron
scattering from polarized $^3$He in the threshold
region for momentum transfers in the range 0--5 fm$^{-1}$.  
These calculations have been based on the Argonne
$v_{18}$ two-nucleon~\cite{WSS95} and
Urbana-IX three-nucleon~\cite{PPCW95}
interactions, and have used accurate bound and
continuum wave functions, obtained with the
PHH method\cite{KRV94,KRV95,KRV93}.  The model for the electromagnetic
operator has been taken to consist of one- and
two-body components, the latter ones constructed
consistently with the two-nucleon interaction~\cite{Car98,Mar98}.
In recent studies, this theory has been shown
to correctly predict the static properties
of the trinucleons~\cite{Mar98} and $A=6$ nuclei~\cite{Wir98}, as well as
their associated elastic and transition
electromagnetic form factors. 

A satisfactory description of all measured $pd$ observables
has emerged with the exception of the differential cross
section and tensor analyzing power at small angles for $E_p<40$ keV,
and the contributions to the GDH integral at energies in the
c.m. range 0--53 keV.  A comparison between the calculated
RMEs and those extracted from fits to the measured data
has shown that the large $p_4$ RME associated with the
channel $L,S,J=1,1/2,3/2$ is overestimated by theory
at very low energy.  It has been speculated that at these
energies (below $\leq 50$ keV) long-range electromagnetic
interactions beyond the static Coulomb repulsion between
protons might play a role.  These electromagnetic
interactions have already been included in bound-state
calculations (such as those reported here),
where they have been found to contribute to
ground-state energy differences between mirror nuclei or
members of isomultiplets~\cite{Pea97}.  However, they
have yet to be included in scattering-state calculations.
Work along these lines is currently underway.  It should
emphasized that at higher energies the splitting
between the $p_2$ and $p_4$ RMEs appears to be much
reduced, as the fits to the $pd$ observables at 2 MeV
indicate.  This result is further corroborated by a
phase shift analysis of $pd$ elastic scattering
at energies below the $ppn$ breakup threshold~\cite{KRTV96}, which
leads to very close values for the $^2P_{1/2}$
and $^2P_{3/2}$ phases. At 2 MeV the tensor observables 
$T_{20}$ and $T_{21}$ are sensitive to the \lq\lq small\rq\rq 
quartet $E_{1}$ RMEs. The \lq\lq full\rq\rq calculation fails 
to correctly predict these observables, suggesting that the 
present model for the electromagnetic transition operator 
may have deficiencies. 
Finally, the validity of the long-wavelength approximation has been analyzed. 
This approximation has been found to be inadequate for the calculation of 
inhibited (and hence \lq\lq small\rq\rq) electric dipole transitions.
Model calculations of low-energy capture and 
photodisintegration observables, based on the long-wavelength 
form of the $E_1$ operator (the LWAc operator of Eq.~(\ref{eq:lw1me})), 
should therefore be viewed with suspicion unless explicitely 
verifying that higher order corrections are indeed negligible.

The body of data available from $^3$He($e,e^\prime$) inclusive
scattering experiments in the threshold region is not as extensive, 
at present, as 
that from $pd$ capture experiments.  However, 
polarized electron scattering data from a polarized
$^3$He target below the $ppn$ threshold
should become available in the near future~\cite{Gao99}.
The longitudinal and transverse data measured at Bates by the Saskatoon
group are in reasonable agreement with theory, although the data
have rather large errors.  The crucial role
played by two-body charge and current operators should be
re-emphasized.  
\section*{Acknowledgments}
The authors wish to thank 
J.L.\ Friar for clarifications concerning
the long-wavelength approximation to electric dipole
transitions, L.D.\ Knutson for useful discussions,  and 
L.D.\ Knutson, D.\ Skopik, H.R.\ Weller and
E.A.\ Wulf for allowing them the use of their data. 
R.S.\ also thanks W.\ Boeglin for a series of interesting
discussions.  M.V.\ and R.S.\ acknowledge partial financial support of NATO
through the Collaborative Research Grant No. 930741.  The support
of the U.S. Department of Energy under contract number DE-AC05-84ER40150
is gratefully acknowledged by L.E.M.\ and R.S.  Finally, some
of the calculations were made possible by grants of time from
the National Energy Research Supercomputer Center in Livermore. 
\appendix
\section{Expressions of observables in terms of reduced matrix elements}
\label{sec:rme}

Expressions for the angular distributions of the
cross section, vector and tensor analyzing powers and 
photon linear polarization coefficient
were derived in terms of reduced matrix elements of
electric and magnetic multipole operators in Ref.~\cite{Viv96}.  It it
useful to extend that analysis also to the case of the
threshold photo- and electro-disintegration of $^3$He, since at the
the small excitation energies of interest here only a relatively
small number of electromagnetic multipoles
are expected to contribute significantly.

The electromagnetic transition amplitudes between an initial
$^3$He bound state with spin projection $\sigma_3$ and
a final $pd$ continuum state having proton and deuteron with
relative momentum ${\bf p}$ and spin projections, respectively,
$\sigma_2$ and $\sigma$, are given by: 

\begin{equation}
   \rho_{\sigma \sigma_2 \sigma_3 }({\bf p},{\bf q})=
  \langle  \Psi_{{\bf p},   \sigma \sigma_2}^{(-)} | \rho({\bf q}) | 
\Psi_{3,{{1\over 2}\sigma_3}} \rangle \ , 
\label{eq:r}
\end{equation}
\begin{equation}
   j^{\lambda}_{\sigma \sigma_2 \sigma_3 }({\bf p},{\bf q})=
  \langle  \Psi_{{\bf p},   \sigma \sigma_2}^{(-)} |
   \hat{\epsilon}_\lambda({\bf q}) \cdot {\bf j}({\bf q}) | 
  \Psi_{3,{{1\over 2}\sigma_3}} \rangle \ ,  
\label{eq:j}
\end{equation}
where ${\bf q}$ is the momentum transfer and $\hat \epsilon_\lambda({\bf q})$,
$\lambda=\pm 1$, are the transverse polarizations of the (real or
virtual) photon. 
The wave function with ingoing-wave boundary condition is expanded as 

\begin{equation}
   \Psi_{{\bf p},\sigma \sigma_2}^{(-)} = 4\pi
   \sum_{SS_z} \langle \frac {1}{2}\sigma,1\sigma_2 |SS_z\rangle
   \sum_{LL_zJJ_z} {\rm i}^L\>
    \langle SS_z,LL_z | JJ_z \rangle Y_{LL_z}^*({\hat {\bf p}})
   {\overline \Psi}_{1+2}^{LSJJ_z(-)}\ ,
\label{eq:psim}
\end{equation}
where the ${\overline \Psi}_{1+2}^{LSJJ_z(-)}$ are related to the
$\Psi_{1+2}^{LSJJ_z}$ introduced in Sec.~\ref{sec:phh} via

\begin{equation}
   {\overline \Psi}_{1+2}^{LSJJ_z(-)}= {\rm e}^{ -{\rm i} \sigma_{L}}
   \sum_{L^\prime S^\prime} \left[ 1+{\rm i}\> { R}^J
    \right]^{-1}_{LS,L^\prime S^\prime}
    \Psi_{1+2}^{L^\prime S^\prime JJ_z }\ .
\label{eq:psipa}
\end{equation}
Here $\sigma_L$ is the Coulomb phase shift and $R^J$
is the $R$-matrix.  Introducing the
expansion above into the matrix elements, Eqs.~(\ref{eq:r})--(\ref{eq:j}),
one finds:

\begin{equation}
j^{\lambda}_{\sigma \sigma_2 \sigma_3 }({\bf p},{\bf q})=4 \pi
\sum_{LL_zSS_zJJ_z} (-{\rm i})^L
\langle \frac {1}{2}\sigma,1\sigma_2 |SS_z\rangle
\langle SS_z,LL_z | JJ_z \rangle Y_{LL_z}(\hat{\bf p})
\> j^{LSJ}_{J_z \lambda \sigma_3}({\bf q}) \>\>,
\end{equation}
\begin{equation}
j^{LSJ}_{J_z \lambda \sigma_3}({\bf q})=
\langle {\overline \Psi}_{1+2}^{LSJJ_z(-)} |
\hat{\epsilon}_\lambda({\bf q}) \cdot {\bf j}({\bf q}) | 
\Psi_{3,{{1\over 2}\sigma_3}} \rangle \ ,  
\end{equation}
and similar expressions hold for the $\rho_{\sigma \sigma_2 \sigma_3}({\bf p},{\bf q})$
amplitudes.  It is now convenient to take $\hat{\bf q}$ as
defining the $z$-axis, i.e. the spin-quantization axis.  Standard
techniques~\cite{Wal95} then lead to the following expansions
in terms of reduced matrix elements of Coulomb ($C$),
electric ($E$) and magnetic ($M$) multipoles for
the amplitudes $\rho^{LSJ}_{J_z \sigma_3}({\bf q})$ 
and $j^{LSJ}_{J_z \lambda \sigma_3}({\bf q})$:

\begin{equation}
\rho^{LSJ}_{J_z \sigma_3}(q \hat{\bf z})=\sqrt{4 \pi}
\sum_{\ell=0}^\infty{\rm i}^{\ell} \sqrt{\frac{2\ell+1} {2J+1}}
\langle \frac {1}{2}\sigma_3,\ell 0 |JJ_z\rangle \>
C^{LSJ}_{\ell}(q) \>\>,
\label{eq:rrme}
\end{equation}
\begin{equation}
j^{LSJ}_{J_z \lambda \sigma_3}(q \hat{\bf z})=-\sqrt{2 \pi} 
\sum_{\ell=1}^\infty {\rm i}^{\ell} \sqrt{\frac{2\ell+1}{2J+1}}  
\langle \frac {1}{2}\sigma_3,\ell \lambda |JJ_z\rangle \>
\lbrack \lambda M_{\ell}^{LSJ}(q) + E_{\ell}^{LSJ}(q) \rbrack  \>\>.
\label{eq:jrme}
\end{equation}

The matrix elements $\rho^{LSJ}_{J_z \sigma_3}(q \hat{\bf z})$
and $j^{LSJ}_{J_z \lambda \sigma_3}(q \hat{\bf z})$ are calculated
with the Monte Carlo integration techniques discussed in Ref.~\cite{Viv96}, and
from these the reduced matrix elements $C^{LSJ}_{\ell}(q)$,
$M_{\ell}^{LSJ}(q)$ and $E_{\ell}^{LSJ}(q)$ are obtained via
Eqs.~(\ref{eq:rrme})--(\ref{eq:jrme}), for example

\begin{equation}
\label{eq:rmee}
  C_0^{LS{1\over2}}(q) =\frac{1}{\sqrt{2 \pi}}
  \rho_{{1\over2} \; {1\over2}}^{LS{1\over2}}(q\hat{\bf z}) \>\>\>, 
\end{equation}
\begin{equation}
  M_1^{LS{3\over2}}(q)= {{\rm i}\over \sqrt{2\pi}}
     \biggl[ \sqrt{3} j_{{3\over2} \;1\;
     {1\over2}}^{LS{3\over2}}(q\hat{\bf z})  -
     j_{-{1\over2} \;-1\; {1\over2}}^{LS{3\over2}}(q\hat{\bf z})
    \biggr]\>\> .
\end{equation}

The inclusive cross section for polarized electron scattering
from a polarized spin 1/2 target can simply be written as~\cite{Don86}

\begin{equation}
\label{eq:sez}
 {d^3 \sigma \over d\Omega d\omega} = \Sigma(q,\omega) + h\> \Delta(q,\omega) \ ,
\end{equation}
\begin{equation}
 \Sigma(q,\omega)= \sigma_M \lbrack v_L R_L(q,\omega) +
 v_T R_T(q,\omega) \rbrack \ , \label{eq:unp} 
\end{equation}
\begin{equation}
 \Delta(q,\omega)= \sigma_M \lbrack v_{LT'} R_{LT'}(q,\omega) \sin\theta^* \cos\phi^* +
                     v_{T'} R_{T'}(q,\omega) \cos\theta^* \rbrack \ , \label{eq:pol}
\end{equation}
where $\sigma_M$ is the Mott cross section,
the coefficients $v_\alpha$ are functions of the electron kinematic
variables, $h=\pm 1$ is the helicity of the incident electron, and the
angles $\theta^*$ and $\phi^*$ specify the direction of the target
polarization with respect to $\hat{\bf q}$.  The response functions $R_\alpha$
contain the nuclear structure information.  We note that  
the sum over the three-nucleon final states, implicit in their
definition, is restricted to include only the
$p$$d$ continuum, since the excitation energies of interest
here are below the threshold for the three-body breakup. 
The longitudinal-transverse and transverse-transverse asymmetries 
$A_{LT^\prime}$ and $A_{T^\prime}$ are defined as:
\begin{eqnarray}
A_{LT^\prime}(q,\omega) &=& {v_{LT^\prime} R_{LT^\prime}(q,\omega) \over 
{v_{L} R_{L}(q,\omega) + v_{T} R_{T}(q,\omega)}} \ ,
\nonumber \\
A_{T^\prime}(q,\omega) &=& {v_{T^\prime} R_{T^\prime}(q,\omega) \over 
{v_{L} R_{L}(q,\omega) + v_{T} R_{T}(q,\omega)}} \ .
\label{eq:asyms}
\end{eqnarray}

Explicit expressions for the response functions $R_\alpha$
in terms of reduced matrix elements of electromagnetic multipole
operators are easily obtained: 

\begin{equation}
R_L=  f_{pd} \sum_{LSJ\ell} |C^{LSJ}_\ell|^2 \ , 
\end{equation}
\begin{equation}
R_T=  f_{pd} \sum_{LSJ\ell}(|E^{LSJ}_\ell|^2 + |M^{LSJ}_\ell|^2 ) \ , 
\end{equation}
\begin{eqnarray}
  R_{LT'}= 2 \sqrt{2}\, f_{pd}
       \sum_{LSJ} \frac{\sqrt{J\!+\!1/2}}{2J\!+\!1}\,
  \Re\Bigg[\Big(\,C^{LSJ}_{-}+{\rm i}\, C^{LSJ}_{+}\,\Big)^*
  \Big[&&\sqrt{J\!-\!1/2}\,(\,M^{LSJ}_{-}+E^{LSJ}_{-}\,) \nonumber \\
&&-{\rm i}\, \sqrt{J\!+\!3/2}\,(\,M^{LSJ}_{+}+E^{LSJ}_{+}\,)\Big] \Bigg] \ ,
\end{eqnarray}
\begin{eqnarray}
R_{T'}&=& 2 \, f_{pd}
       \sum_{LSJ} \frac{1}{2J+1}
\Bigg[|M^{LSJ}_{-}+E^{LSJ}_{-}|^2-|M^{LSJ}_{+}+E^{LSJ}_{+}|^2 \nonumber \\
&&-2\, \sqrt{(J\!+\!3/2)(J\!-\!1/2)}\, \Im\, \Big[ (\,M^{LSJ}_{-}+E^{LSJ}_{-}\,)^* 
                                 (\,M^{LSJ}_{+}+E^{LSJ}_{+}\,)\Big] \Bigg] \ ,
\end{eqnarray}
where the phase-space factor $f_{pd}$ is given by $f_{pd}=4 \mu  p$, and
in the interference response functions the notation
$X^{LSJ}_{\pm}$ for the reduced matrix elements
means $X^{LSJ}_{\ell=J \pm 1/2}$.  The
magnitude of the relative momentum $p$ is fixed by energy conservation

\begin{equation}
\omega+E_3=E_2+\frac{q^2}{2(m_2+m)}+\frac{p^2}{2 \mu} \>\>,
\end{equation}
where $E_2$ and $E_3$ are the two- and three-body ground-state energies,
$m_2$ is the deuteron mass and $\mu$ is the 1+2 reduced mass.

The photo-disintegration cross section is simply related to
the $R_T$ response function:

\begin{equation}
\sigma^\gamma(\omega) = \frac{4 \pi^2 \alpha}{\omega} R_T(\omega) \>\>,
\end{equation}
where for real photons $q =\omega$.  In particular,
the difference of cross sections in the integrand 
of Eq.(\ref{gdh}) is easily related to the
response function $R_{T'}$
\begin{equation}
\sigma^\gamma_P(\omega) -\sigma^\gamma_A(\omega)= 
\frac{4 \pi^2 \alpha}{\omega} R_{T'}(\omega) \>\>.
\end{equation}

%
%
%
\begin{table}
\begin{tabular}{ldddd}
&\multicolumn{2}{c}{$a_2$ (fm)}&\multicolumn{2}{c}{$a_4$ (fm)}\\
      & Th. & Exp. & Th. & Exp. \\
\tableline
$nd$   &  0.63   & 0.65 $\pm$ 0.04  & 6.33  & 6.35 $\pm$ 0.02   \\
$pd$   &--0.02   &                  & 13.7      &    \\
\end{tabular}
\caption{Predictions obtained from the AV18/UIX Hamiltonian
model with the PHH method for the $nd$ and $pd$ doublet and
quartet scattering lengths $a_2$ and $a_4$.}
\label{tab:lgths}
\end{table}
\begin{table}
\begin{tabular}{lddddd}
& IA & FULL & LWAc & LWAb & LWA2 \\
\tableline
 $\widetilde E_{1C}^{1{1\over2}{1\over2}}$ &
--24.2& --26.2 & --26.0& --0.1 & 0.1 \\
 $\widetilde E_{1C}^{1{3\over2}{1\over2}}$ & 
--6.1 &  0.6   &   2.1 &   0.6 & --0.4 \\
 $\widetilde E_{1C}^{1{1\over2}{3\over2}}$ & 
33.1  &  36.7  &  36.9 &   0.0 & 0.1 \\
 $\widetilde E_{1C}^{1{3\over2}{3\over2}}$ & 
--2.7 &  0.1   & 0.6   & --1.3 & 0.7
\end{tabular}
\caption{
Doublet and quartet $E_1$ RMEs in fm$^{3/2}$ calculated with the
AV18/UIX Hamiltonian model for the reaction
$^2$H($p$,$\gamma$)$^3$He at zero energy in IA and FULL 
approximations. The contributions LWAc, LWAb, and LWA2 are
reported in columns 4--6, see text for an explanation.  Note that
the LWA2 contribution has been calculated by using only the magnetization
term of the single-nucleon current, i.e. by only retaining 
the second term on the r.h.s. of Eq.~(\ref{eq9}). 
Statistical errors associated with the Monte Carlo integrations are in
the range 1--5\%. 
}
\label{tab:lwa}
\end{table}

\begin{table}
\begin{tabular}{lddd}
RME & IA & FULL & FIT \\
\tableline
 $|\tilde m_2|$ & 0.172  & 0.322 & 0.340$\pm$0.010  \\
 $|\tilde m_4|$ & 0.174  & 0.157 & 0.157$\pm$0.007  \\
 $|\tilde p_2|$ & 0.346  & 0.371 & 0.363$\pm$0.014  \\
 $|\tilde p_4|$ & 0.343  & 0.378 & 0.312$\pm$0.009  \\
\end{tabular}
\caption{Magnitudes of the leading $M_1$ and $E_1$ RMEs 
for $pd$ capture at $E_p=40$ keV.  The values listed in the fourth column
have been determined in Ref.~\protect\cite{Wea99} from a fit to
the measured observables, while those listed in the second and third
columns have been obtained in calculations using either one-body (IA) only
or both one- and two-body (FULL) currents.  Note that the RMEs listed in
this table are adimensional.}
\label{tab:rmew}
\end{table}

\begin{table}
\begin{tabular}{ddd}
$E_p$(MeV) & ${}^2P_{1/2}$ &  ${}^2P_{3/2}$ \\
\tableline
0.035&  --0.00392 &    --0.00391 \\
0.213& --0.699   &    --0.695   \\
2.0  &  --4.89 &     --4.84  \\
3.0  &  --7.37   &    --7.17  \\
\tableline
3.0 (PSA)  &  --7.41(0.08) & --7.18(0.04) \\
\end{tabular}
\caption{Nuclear elastic $pd$ phase shifts (in degrees) obtained for a few
selected proton energies with the AV18/UIX Hamiltonian model.
The values extracted from the phase-shift
analysis (PSA) of Ref.~\protect\cite{KRTV96} are listed in the last row.}
\label{tab:epd}
\end{table}

\begin{table}
\begin{tabular}{cddd}
RME & set 1 &  set 2 & AV18/UIX  \\
\tableline
$\overline {\cal M}_1^{0{1\over 2}{1\over 2}}$ & --0.221 &         & --0.243\\
$\overline {\cal M}_1^{2{3\over 2}{1\over 2}}$ &        & --0.355  & 0.023\\
\tableline
$\overline {\cal M}_1^{0{3\over 2}{3\over 2}}$ & --0.221 & --0.220  & --0.192 \\
$\overline {\cal M}_1^{2{1\over 2}{3\over 2}}$ &  0.272 &  0.476  & 0.018 \\
$\overline {\cal M}_1^{2{3\over 2}{3\over 2}}$ &  0.184 &  0.324  & 0.019 \\
\tableline
$\overline {\cal E}_1^{1{1\over 2}{1\over 2}}$ & 2.721  & 2.434  & 2.678  \\
$\overline {\cal E}_1^{1{3\over 2}{1\over 2}}$ & --0.122 & --0.118  & --0.308\\
\tableline
$\overline {\cal E}_1^{1{1\over 2}{3\over 2}}$ & 2.742 & 2.717  & 2.771 \\
$\overline {\cal E}_1^{1{3\over 2}{3\over 2}}$ & 0.080 & 0.061  & 0.175 \\
$\overline {\cal E}_1^{3{3\over 2}{3\over 2}}$ & 0.061 & 0.085  & 0.072 \\
\tableline
\end{tabular}
\caption{Real RMEs ($\times 10^3$) for $pd$ radiative capture 
at $E_{\rm c.m.}=2$ MeV, see text for definitions.  Sets labelled
1 and 2 are the RMEs obtained in Ref.~\protect\cite{SK99} from a
fit to the measured observables.  The RMEs in the
column labelled AV18/UIX have been obtained by the calculation
presented in this paper, including the full current
operator.  Note that the RMEs defined in this table are adimensional.}
\label{tab:rmeP}
\end{table}
\begin{table}
\begin{tabular}{cdddddd}
RME & set 1 &  IA & FULL & LWAc & LWAb & LWA2 \\
\tableline
$\overline {\cal E}_1^{1{1\over 2}{1\over 2}}$ & 
  2.721   & 2.352   & 2.678    & 2.699  & 0.012  & --0.021  \\
$\overline {\cal E}_1^{1{3\over 2}{1\over 2}}$ & 
  --0.122 & --1.166 & --0.308  &--0.127 & 0.071  & --0.050 \\
\tableline
$\overline {\cal E}_1^{1{1\over 2}{3\over 2}}$ & 
    2.742 & 2.395   & 2.771    & 2.716  & --0.042& 0.011 \\
$\overline {\cal E}_1^{1{3\over 2}{3\over 2}}$ & 
    0.080 & 0.405   & 0.175    & 0.111  & 0.121  & --0.069 \\
$\overline {\cal E}_1^{3{3\over 2}{3\over 2}}$ & 
    0.061 & 0.069   & 0.072    & 0.070  & 0.002  & 0.001 \\
\tableline
\end{tabular}
\caption{Real RMEs ($\times 10^3$) for $pd$ radiative capture 
at $E_{\rm c.m.}=2$ MeV as in Table~\ref{tab:rmeP}.  Set 1 
are the RMEs obtained in Ref.~\protect\cite{SK99} from a
fit to the measured observables.  The RMEs in the
columns labelled IA, FULL, LWAc, LWAb, and LWA2 have been
obtained in calculations using various approximations
for the $E_1$ operator as in Table~\ref{tab:lwa}.
Note that the RMEs defined in this table are adimensional.} 
\label{tab:lwP}
\end{table}
\begin{table}
\begin{tabular}{cdd}
 wave & $\delta^{LSJ}$ Exp &  $\delta^{LSJ}$ AV18/UIX \\
        & (deg) & (deg) \\
\tableline
 ${}^2S_{1/2}$ & --24.9  & --27.9 \\
 ${}^4D_{1/2}$ &    4.27 &    4.27\\
\tableline
 ${}^4S_{3/2}$ & 116.1   & 116.9 \\
 ${}^2D_{3/2}$ &    9.83 &   9.98\\
 ${}^4D_{3/2}$ &    4.12 &   4.01\\
\tableline
 ${}^2P_{1/2}$ & --1.82 & --2.15 \\
 ${}^4P_{1/2}$ &  26.9  &  27.4 \\
\tableline
 ${}^2P_{3/2}$ & --1.95 & --1.95\\
 ${}^4P_{3/2}$ &  29.4  &  29.3 \\
 ${}^4F_{3/2}$ &  10.4  &  10.4 \\
\tableline
\end{tabular}
\caption{Eigenphase shifts $\delta^{LSJ}$ for $pd$ elastic scattering 
at $E_{\rm c.m.}=2$ MeV.  The values of $LSJ$ are given in column 1 in
the format ${}^{2S+1}L_J$.  The experimental phase shifts are those
from Ref.~\protect\cite{SK99}, whereas the theoretical ones are calculated
with the AV18/UIX interaction model.  The values
reported here are the phase shifts induced in the wave functions by
the nuclear plus Coulomb potential.
}
\label{tab:eps}
\end{table}
\begin{table}
\begin{tabular}{ccccc}
$\overline{\omega}$ (MeV) &  $I(\overline{\omega})$     & FIT     & IA & FULL\\
\hline
5.522 &  $M_1$         & $-0.0524\pm 0.0077$ & $-0.0029$ & $-0.0609$ \\
5.522 &  $E_1$ $S=$1/2 & $-0.0361\pm 0.0095$ & $-0.0030$ & $+0.0027$ \\
5.522 &  $E_1$ $S=$3/2 & $-0.0026\pm 0.0007$ & $-0.0050$ & $-0.0001$ \\
\hline
5.522 &  Total     & $-0.0911\pm 0.0123$ & $-0.0109$ & $-0.0583$ \\
\hline \hline
5.548 &  Total    & $-1.120 \pm 0.218 $ & $-0.161 $ & $-0.582 $
\end{tabular}
\caption{The contributions $I(\overline{\omega})$ (in nb) for two
energies $\overline{\omega}$.  In the third column, the
values obtained from a fit of the
experimental $pd$ capture data are reported~\protect\cite{Wea99}.
The results of the theoretical calculations using the AV18/UIX Hamiltonian
model and either one-body only or both one- and two-body currents
are listed in the fourth and fifth columns, labelled IA and FULL
respectively.  The lines denoted by $M_1$, $E_1$ $S$=1/2 and $E_1$ 
$S$=3/2 report
the partial contributions to $I(\overline{\omega}$=5.522 MeV$)$ of the
corresponding RMEs.}
\label{tab:res}
\end{table}
%
%
\begin{figure}[p]
\caption{The $S$-factor for the ${}^2$H($p$,$\gamma$)${}^3$He reaction,
obtained with the AV18/UIX Hamiltonian model and one-body only (dashed line)
or both one- and two-body (solid line) currents,
is compared with the experimental values of Refs.~\protect\cite{Sea96,Mea97}.}
\label{fig:S}
\end{figure}

\begin{figure}[p]
\caption{The energy integrated cross section $\sigma(\theta)/a_0$
($4\pi a_0$ is the total cross section), vector analyzing power
$A_y(\theta)$, tensor analyzing power $T_{20}(\theta)$ and photon
linear polarization coefficient $P_\gamma(\theta)$ obtained with the
AV18/UIX Hamiltonian model and one-body only (dashed line)
or both one- and two-body (solid line) currents 
are compared with the experimental results of Ref.~\protect\cite{Sea96}.}
\label{fig:cpt080keV}
\end{figure}

\begin{figure}[p]
\caption{The functions $\Re[\tilde p_{2J+1}(r_{pd})]$ calculated with the
AV18/UIX Hamiltonian model for the states $J=1/2$ and $3/2$.  The functions
obtained by switching off the nuclear $pd$ intercluster interactions
are displayed by the thin dashed and solid lines for the
$J=1/2$ and $J=3/2$ scattering states, respectively.  The two lines are
indistinguishable.  When the nuclear interactions between the $d$ and
$p$ clusters are taken into account, the functions are shown by the thick 
dashed and solid lines for the $J=1/2$ and $J=3/2$ scattering states,
respectively.}
\label{fig:prd}
\end{figure}

\begin{figure}[p]
\caption{Proton vector analyzing power $A_y$ and
deuteron tensor analyzing power $T_{20}$ for $pd$ capture at 
$E_{\rm c.m.}=75$ and $100$ keV,
obtained with the AV18/UIX Hamiltonian model 
and one-body only (dashed lines)
or both one- and two-body currents (solid lines).
The experimental values are from Ref.~\protect\cite{Mea97}.}
\label{fig:ayt20}
\end{figure}

\begin{figure}[p]
\caption{Differential cross section, proton vector analyzing power, and the
four deuteron tensor analyzing powers for $pd$ capture at $E_{\rm c.m.}=2$ MeV,
obtained with the AV18/UIX Hamiltonian model 
and one-body only (dashed lines)
or both one- and two-body currents (thin solid lines), 
are compared with the
experimental values of Ref.~\protect\cite{SK99}.  The results
obtained in the approximation LWAc for the $E_1$ operator are
also shown (solid lines).}
\label{fig:cpt2MeV}
\end{figure}

\begin{figure}[p]
\caption{The $S$-factor for the ${}^2$H($p$,$\gamma$)${}^3$He
reaction, in the c.m. energy range $0-2$ MeV,
obtained with the AV18/UIX Hamiltonian model and
one- and two-body currents (solid line)
is compared with the experimental values listed in the web site
{\tt http://pntpm.ulb.ac.be/nacre.htm}.}
\label{fig:S2}
\end{figure}

\begin{figure}[p]
\caption{The longitudinal and transverse response functions of $^3$He,
obtained with the AV18/UIX Hamiltonian model and one-body only
(dashed lines) or both one- and two-body (solid lines) charge and current
operators, are compared with the data of Ref.~\protect\cite{Rea94} at excitation 
energies below the $ppn$ breakup threshold.}
\label{fig:res3he}
\end{figure}

\begin{figure}[p]
\caption{The longitudinal ($R_L$) and longitudinal-transverse 
($R_{LT^\prime}$) response functions of $^3$He,
obtained with the AV18/UIX Hamiltonian model and one-body only
(thick dashed lines) or both one- and two-body (thick solid lines)
charge and current operators,
are displayed at a fixed excitation 
energy of 1 MeV for three-momentum transfers in the range 0--5 fm$^{-1}$.
The contributions associated with the (dominant) S-wave $pd$ scattering states
are also shown.}
\label{fig:rl-rlt}
\end{figure}

\begin{figure}[p]
\caption{Same as in Fig.~\protect\ref{fig:rl-rlt}, but for the
transverse ($R_T$) and transverse-transverse ($R_{T^\prime}$)
response functions of $^3$He.  The contributions associated with
both S- and P-wave $pd$ scattering states are also shown.}
\label{fig:rt-rtt}
\end{figure}

\end{document}